\documentstyle[preprint,aps,prb]{revtex}
\newcommand{\cn}[1]{\frenchspacing\onlinecite{#1}}
\begin{document}
\draft
\title{Wavy film flows down an inclined plane. 
Part I: Perturbation theory and 
general evolution equation for the film thickness}
\author{A. L. Frenkel and K. Indireshkumar}
\address{Department of Mathematics,\\
University of Alabama,\\
Tuscaloosa, Alabama 35487-0350}
\date{\today }
\maketitle

\begin{abstract}
Wavy film flow of incompressible Newtonian fluid down an inclined plane
is considered. The question is posed as to the parametric conditions
under which the description of evolution can be approximately reduced
for all time to a single evolution equation for the film thickness. 
An unconventional perturbation approach yields the most general evolution
equation and least restrictive conditions on its validity. The advantages
of this equation for analytical and numerical studies of 3D waves  
in inclined films are pointed out.
\end{abstract}
\preprint{HEP/123-qed} \narrowtext
\section{INTRODUCTION}
\label{s1} Thin liquid layers (``films'') flowing along solid
surfaces---such as inclined planes or vertical cylinders---occur in both
natural and man-made environments. Industrial applications of film flows
started as long ago as the
1800s and have been growing in their scope and
importance ever since (see e.g. Refs. \cn{f64,an94}). 
 
Accordingly, the studies of film flows, in particular those down an inclined
plane, have a considerable history (see e.g. Refs. \cn{f64,lp93}). 
However, the dynamics of nonlinear waves (typically present in film flows)
is still far from being satisfactorily understood (see Refs. \cn{fi96,cd96}  
for the most recent 
progress reviews). The nonlinear Navier-Stokes
(NS) partial differential equations (PDEs) of this ``Kapitza problem''
couple together several fields (pressure and the components of velocity),
each a function of time and three spatial coordinates. Furthermore, the
boundary conditions (BCs) of the problem involve a free boundary (the free
surface of the film) whose PDE itself is coupled to the NS equations. The
full spatially 3-dimensional (3D) problem is too hard to simulate even with
the most powerful modern computers.
 
Even simpler, 2D computations of wavy films have been undertaken only under
the simplifying assumptions of short spatial intervals (e.g. 
Refs. \cn{jd92a,cr94}) and/or 
time-independence (e.g. Ref. \cn{sa94}). However, 2D flows
are frequently unstable to 3D disturbances, and three-dimensionality can be
important for many inclined films (see e.g. recent experiments 
in Ref. \cn{ls95}).
 
Therefore, naturally, one looks for more manageable {\it approximate}
descriptions of wavy-film evolution. Such simplified theories are possible
in certain domains of the space of parameters (for which, typically, the
slopes of the surface waves are small). Of course, the greater
simplification has to be paid for by a more limited applicability of the
theory. The greatest simplification is achieved when the problem reduces to
a single evolution PDE approximating the thickness of the film. [From the
numerical-simulation point of view, the most important simplification
here is the reduction in the number of independent spatial variables,
and even a lower-dimensional {\it system} of equations would have been
not much more difficult than a {\it single} evolution equation (EE).
However, no such system has ever been derived consistently for an inclined
single layer film, the subject of our consideration in this paper. The
same is true for the {\it nonlocal}, i.e. integro-differential,
film-thickness equations. (The only known example\cite{f88} of such a
nonlocal EE, as far as film flows are concerned, is an EE for a 
core-annular flow, and not for an inclined film.) In any case, in this
paper, the nonlocal equations are altogether excluded from the consideration;
thus, when we say ``the most general EE'' it should be understood as
``the most general {\it local} EE'', etc.]
 Although
less drastic simplifications than a single EE, 
and therefore having a larger parametric range of validity,
are known (see e.g. Ref. 
\cn{cd96}), so far fully-dimensional simulations for
sufficiently extended spatial domains have been carried out\cite{fi96,do91}
only for the theories hinged on a single evolution equation (EE). Such
single-EE theories of inclined-film flows are the subject of the present
paper.

Evolution equations for film thickness have been known since the pioneering
work of Benney.\cite{b66} The conventional perturbation approach to their
derivation (e.g. Refs.
\cn{g70,l69,n75,r70}) used a small (longwave) parameter,
say $\epsilon $. In particular, each of the ``global'' (``internal'',
``basic'') parameters specifying the problem (the parameters appearing in
the dimensionless NS equations and the free-surface BCs) must be ascribed,
in such a single-parameter (SP) technique, a
certain power of $\epsilon $ as its order of magnitude ($O_M$). Therefore,
an artificial dependence is forced on the---intrinsically
independent---parameters. This unnecessarily restricts the domain of
justified validity of the resulting EE. For example, for the vertical film,
there are just two independent parameters: the ``Reynolds number'' $R$ and
the ``Weber number'' $W$. If $R$ is of the order of magnitude of $(\sim )$ $%
\epsilon ^a$ and $W$ $\sim $ $\epsilon ^b$, then $W\sim R^{b/a}$. The set of
points $W=R^{b/a}$ is just a 1D curve in the 2D space ($R$, $W$), while the
complete domain for which the EE is valid is likely to have the same
dimensionality as the parameter space ($R$, $W$) itself, that is it should
be a 2D domain.

Furthermore, each time the powers assigned 
to the parameters are altered, it is in principle necessary
to again go through the entire procedure or the SP derivation, and
one can
arrive at a different EE as a result. 
For example, Topper and Kawahara \cite{tk78}
considered two cases of an inclined-film flow (such a flow is specified by 
{\it three} global parameters: in addition to $R$ and $W,$ the inclination
angle $\theta $ can be independently varied). In one case, they required the
angle $\phi$ $(\equiv \pi /2-\theta )$ of the film plane with the vertical
to be small, $\phi \sim \epsilon ,$ and stipulated $R\sim \epsilon ^1$ and $%
W\sim \epsilon ^{-1}$ (in our definitions of $R$ and $W$; see section \ref
{s2} below). As a result, they obtained an EE containing {\it both}
dissipative and dispersive terms (and with all coefficients in the EE being $%
\sim $ 1). However, their second case, for which they chose $\cot \theta
\sim 1$, $R\sim 1$ and $W\sim \epsilon ^{-1}$, resulted in an equation with 
{\it no} dispersive terms. (In the SP framework, 
it is only formally that the latter equation can
be obtained from the former by omitting its dispersive term, and the only
way to really ``justify'' the nondispersive equation is to repeat the
entire derivation procedure starting all the way back from before
NS equations.)

If a given inclined-film system is not close to any of these two
parameter-space curves, the theory \cite{tk78} is invalid.
Logically, there are three possibilities: (i) the flow evolution
cannot be (approximately) reduced to a single EE for all time; (ii) such a
reduction {\it is} possible but the resulting EE is 
different from each of those obtained in 
Ref. \cite{tk78}; or (iii) the EE {\it coincides} 
with one of their two EEs, despite
the different parameter curve. We are naturally led to the following
questions: (i) under what (parametric) conditions 
is an approximate description
of the film flow possible (for all time) which can be reduced
to a single EE? (ii)
How can the set of all such EEs be characterized? These are the questions we
pose and attempt to answer, for inclined-film flow, in this paper. 

We note that one must distinguish between the {\it all-time} validity
of an EE and the {\it limited} in time validity. This
issue arises because in the SP approach, the fixed powers of the small 
parameter are stipulated not only for the global parameters of the 
system, which do not depend on time, but also for the characteristic
time- and length-scales. However, these characteristic scales can change
with time as the long---dissipative---system proceeds to the attractor;
they are instantaneous parameters. So, the assumption
of fixed scales is incorrect after a limited time has elapsed, and 
therefore, the SP derivation is invalid.

The rest of the paper is organized as follows.
In section II, we formulate the full NS problem. In section III, we
introduce an iterative perturbation procedure. It starts with the well-known
(although typically unstable), waveless, ``Nusselt'' solution of the NS
problem. The only principle necessary in deciding the iteration steps is
the requirement that in the end, a {\it single} 
(and valid for all time) EE should be arrived
at, with minimal simplification of exact equations. 
No dependencies are imposed on the internal parameters: the validity
conditions (VCs) we obtain require that several quantities, which are
certain products of powers of the internal parameters, be small---{\it %
independently} of one another. [So, {\it several} independent (small)
parameters emerge in the derivation. Thus, the iterative procedure we
introduce here is a variation of the ``multiparameter'' perturbation (MP)
approach developed in Refs. \cn{fb87,f88,f91,f92}.]

In section IV, we arrive at the most general evolution equation
(GEE) valid (provided 
certain restrictions on parameters are satisfied) for all time: 
Any all-time-valid EE 
derivable with a conventional single-parameter (SP) approach 
necessarily coincides with one of just a few prototype equations
which are certain truncations of the GEE.
The GEE also has the least restrictive domain of
validity: It is easy to obtain the domain of
validity for each simplified EE, and it is a subdomain of the all-embracing
(corresponding to the GEE) validity domain. Outside of the latter
parametric domain,
there is {\it no} single EE which could approximate the film evolution for
all time. That this is the case is argued in section V by analyzing the
structure of possible correction terms of the EE through all orders of the
iteration procedure. (In particular, as was shown in Ref.
\cn{f93} and 
Ref. \cn{if95}---where a different 
derivation of the same EE was sketched---the 
{\it amplitudes} of inclined-film waves 
which can be described (for all time) by a single
EE are necessarily {\it small}.) The paper is summarized in the last
section. Some of the more technical considerations 
are relegated to Appendices.

Simulation results for different versions of the inclined-film flow EEs we
obtain here will be given in a sequel 
to this paper. [For some of those results (in
particular, those showing good agreement with experiments \cite{ls95}),
see Ref. \cn{fi96}.]

\section{The exact Navier-Stokes problem}
\label{s2} 
We consider a layer of an incompressible Newtonian liquid flowing
down an inclined plane under the action of gravity. Our (cartesian)
coordinates are as follows: the $\overline{x}$ axis is normal to the plane
and directed into the film; the $\overline{y}$ axis is in the spanwise
direction; and the $\overline{z}$ axis is directed streamwise (the overbar
here and below indicates a {\it dimensional} quantity). The corresponding
components of velocity are $\overline{u}$, $\overline{v}$, and $\overline{w}$%
. We denote by $\overline{p}$ the pressure field in the film; the pressure
of the ambient passive gas is neglected for simplicity.

The system is determined by the following independent (dimensional)
parameters: the average thickness of the film $\overline{h}_0$; the liquid
density $\overline{\rho }$, viscosity $\overline{\mu }$, and surface tension 
$\overline{\sigma }$; gravity acceleration $\overline{g\text{;}}${\it \ }and
the angle of the plane with the horizontal $\theta $.

There is a well-known time-independent, Nusselt's solution of the NS
problem for the inclined film. The thickness of the Nusselt film is constant
(hence, Nusselt's flow is also referred to as a ``flat-film'' solution). The
only nonzero component of velocity is the streamwise one. It only changes
across the film, starting from the zero value at the solid plane. The
free-surface value $\bar U$ of the Nusselt velocity is $\bar U=\bar g\bar h%
_0^2\sin \theta /(2\bar \nu )$ (where $\bar \nu =\overline{\mu }/$ $%
\overline{\rho }$ is the kinematic viscosity). We nondimensionalize all
quantities with units of measurement based on $\overline{\rho }$, $\overline{%
h}_0$, and $\bar U$ (we have, e.g., the following units: $\overline{h}_0$
for all coordinates; $\bar U$ for velocities; $\overline{h}_0/\bar U$ for
time $t$; $\overline{\rho }\bar U^2$ for pressure; 
$\bar{\rho}\bar U^2\bar{h}_0$ for
surface tension; etc.). We will see that exactly three independent {\it %
basic parameters} (BPs){\it \ }appear in the {\it dimensionless} equations
and boundary conditions of the problem; one can choose e.g. the inclination
angle $\theta $, the Reynolds number $R\equiv \bar h_0\bar U/\bar \nu~ [=%
\bar g\bar h_0^3\sin \theta /(2\bar \nu ^2)]$, and the Weber number $W\equiv
\sigma R/2~ [=\overline{\sigma }/(\overline{\rho }\bar g\bar h_0^2\sin \theta
)]$, as such BPs.

We can write the Navier-Stokes momentum
equations in coordinates moving relative
to the solid plane with a constant velocity $V$ in the $z$-direction (i.e.
introducing $\widetilde{z}=z-Vt$ and omitting the tilde) in the following
dimensionless form (see e.g. Ref. \cn{ah76}): 
\[
u_t+uu_x+vu_y+wu_z-Vu_z= 
\]
\begin{equation}
-p_x-\frac 2R\cot \theta+\frac 1R (u_{xx}+u_{yy}+u_{zz}),  \label{e01a}
\end{equation}
\[
v_t+uv_x+vv_y+wv_z-Vv_z= 
\]
\begin{equation}
-p_y+\frac 1R (v_{xx}+v_{yy}+v_{zz}),  \label{e01b}
\end{equation}
\[
w_t+uw_x+vw_y+ww_z-Vw_z= 
\]
\begin{equation}
-p_z+\frac 2R+\frac 1R (w_{xx}+w_{yy}+w_{zz}).  \label{e01c}
\end{equation}
(The subscripts $x$, $y$, $z$, and $t$ here and
below denote the corresponding partial
derivatives. We will see below that it is appropriate to 
choose $V=2$, the common phase velocity of 
all infinitesimally weak waves.) 
The continuity equation is 
\begin{equation}
u_x+v_y+w_z=0.  \label{e02}
\end{equation}
The BCs are as follows. The no-slip conditions at the
solid plane are 
\begin{equation}
u=v=w=0~~~~~~~(x=0).  \label{e03}
\end{equation}
The tangential-stress balance conditions at the free surface $x=h(y,z,t)$,
the local film thickness, are 
\[
(v_x+u_y)(1-h_y^2)+2(u_x-v_y)h_y-(v_z+w_y)h_z 
\]
\begin{equation}
-(u_z+w_x)h_yh_z=0~~~~~(x=h)  \label{e04a}
\end{equation}
and 
\[
(u_z+w_x)(1-h_z^2)+2(u_x-w_z)h_z-(v_z+w_y)h_y 
\]
\begin{equation}
-(u_y+v_x)h_yh_z=0~~~~~(x=h).  \label{e04b}
\end{equation}
The normal-stress balance condition is 
\[
-p(1+h_y^2+h_z^2)^{3/2}-\frac 2R\left[ u_x+v_yh_y^2+w_zh_z^2\right. 
\]
\[
\left. -(u_y+v_x)h_y+(v_z+w_y)h_yh_z\right. 
\]
\[
\left. -(u_z+w_x)h_z\right] \left ( 1+h_y^2+h_z^2\right )^{1/2} 
=\sigma\left  [h_{yy}(1+{h_z}^2)\right .
\]
\begin{equation}
\left .
+h_{zz}(1+{h_y}^2)-2h_yh_zh_{zy}\right ] ~~~~~ (x=h).
\label{e05}
\end{equation}
Finally, the kinematic condition at the free surface is 
\begin{equation}
h_t+vh_y+wh_z-Vh_z=u~~~(x=h).  \label{e08}
\end{equation}

\section{Iterative perturbation procedure}

\label{s3}

\subsection{Minimal requirement of derivability}

As was motivated in the Introduction, we are interested in the question of
(approximate) reducibility of the above complicated description of
inclined-film dynamics to a single EE. It appears that an {\it iterative}
perturbation approach is appropriate for the general analysis of the problem.

Looking at the previously known, conventional (single-parameter expansion)
derivations of EEs (valid each for its own particular curve
in the parameter space; see
the Introduction), it is clear that each of them in effect discards some terms
of the NS momentum equations so that each of those essentially becomes an
ordinary differential equation (ODE) in $x$, linear and with constant
coefficients, that is easy to solve (with similar simplifications in BCs).
When these solutions (for velocities in terms of film thickness) are
substituted into the kinematic condition (\ref{e08}), the EE for thickness $%
h $ (or, equivalently, for the thickness deviation $\eta \equiv h-1$)
ensues. However, after all the quantities have been expanded in power of $%
\epsilon $ (see Introduction), one has no control over which NS terms to
omit: this is simply dictated by the expansion scheme of the SP
approach. Thus, sometimes
``harmless'' terms are discarded: even if they were retained, one still
would be able to solve for velocities, arriving, as a 
result, at a clearly {\it more general} EE.

Accordingly, 
our main idea in this paper is to look for a derivation 
in which the single postulated
requirement would be that a maximally
general EE for film thickness be arrived at in the end;
so only those terms of the exact NS equations will be discarded which are
clearly in the way of obtaining linear ODEs for velocities and pressure [we
call this the ``minimal requirement of
derivability'' (MRD)]. In this way we obtain approximate
solutions for the deviations of exact solutions from the ``seed'' Nusselt's
fields [e.g., the approximation $w_0$ to $\widetilde{w_0}\equiv w-w_N,$
where the Nusselt $w_N$ is known: see (\ref{ewn}) below]. For some parameter
values, this immediately leads to an EE whose approximation of exact
evolution is good for all time (here, as was mentioned in the
Introduction, we are only interested in such
all-time-valid EEs; see e.g. Ref.
\cn{fi96} for a further discussion of their
difference from the EEs whose validity is {\it limited}
in time). In
other cases, however, the procedure must be repeated, with the refined
solutions, such as $w_N+w_0,$ playing the seed role that was played by the
Nusselt solutions on the original stage (so that at the second iteration
stage one determines the approximation $w_1$ to $\widetilde{w_1}\equiv
w-w_N-w_0$, etc.). Thus, ours is an {\it
iterative} perturbation approach---which
is known even in general to be an alternative 
to the {\it expansion} method (see e.g. 
Ref. \cn{h91}).

Estimating all members of NS problem equations in terms of parameters, the
requirement that the discarded members be much smaller than those retained
leads to the parametric
conditions for our derivation to be valid (see
Appendix \ref{aa1}). Thus, the method yields (i) the evolution equation for
film thickness; (ii) explicit expressions for velocity components and
pressure in terms of the film thickness; and (iii) the parametric validity
conditions of the theory.

It is known (see Ref. \cn{fi96} and references therein) that no single-EE
description can exist globally in time for those parametric regimes of
inclined-film flow which lead to the eventual
amplitude of surface waves being
``large''---comparable to the average film thickness. But in the present
communication, as was mentioned above, we are interested exactly in the {\it %
large-time} behavior, when the system is already close to the attractor, and
we want a {\it single-EE} description of the wavy film dynamics. Therefore,
in addition to the MRD, we can use from the very
beginning---in order to simplify our derivation---the requirement that the
amplitude $A$ of the film thickness deviation, $A(t)\equiv \max \left| \eta
\right| ,$ be small (for all time): with 
\begin{equation}
h=1+\eta ,  \label{eh0}
\end{equation}
we have 
\begin{equation}
\mbox{max}|\eta (y,z,t)|=A\ll 1.  \label{evc0}
\end{equation}
[Note that $A$ can depend upon time; in such cases, we say that the
parameter is a {\it local parameter}, in contrast to the time-independent,
``global'' BPs $\theta $, $R$, and $W$.] However, unlike the conventional
derivations, we do not have to postulate that the characteristic
lengthscales in the film plane are large (the longwave assumption); rather,
this will {\it follow} from the MRD.

\subsection{Nusselt's solution}

The above-mentioned Nusselt solution (of the NS problem) which is steady
and uniform along the film (i.e. in the streamwise and spanwise directions)
is as follows. The dimensionless Nusselt streamwise velocity $w_N$ is 
\begin{equation}
w_N(x)=2x-x^2.  \label{ewn}
\end{equation}
This clearly satisfies the $z$-NS equation 
\begin{equation}
w_{Nxx}=-2  \label{edewn}
\end{equation}
with the boundary conditions 
\begin{equation}
w_{Nx}=0~~~(x=1)  \label{ebcwn0}
\end{equation}
and 
\begin{equation}
w_N=0~~~(x=0).  \label{enswn}
\end{equation} 
 
Similarly, the Nusselt pressure 
\begin{equation}
p_N=\frac 2R (\cot \theta )(1-x)  \label{epn}
\end{equation}
is the solution of the $x$-NS equation 
\begin{equation}
p_{Nx}=-\frac 2R \cot \theta  \label{edepn}
\end{equation}
with the boundary condition 
\begin{equation}
p_N=0~~~(x=1).  \label{ebcpn}
\end{equation}
Finally, the Nusselt normal and spanwise velocities are 
\begin{equation}
u_N=0~~\mbox{ and }~~~v_N=0.  \label{evun}
\end{equation}

At sufficiently large Reynolds number, the destabilizing effect of inertia
overcomes the stabilizing influence of gravity so that the Nusselt solution
loses its stability. (An analysis later in this section leads to $%
R>(5/4)\cot \theta$, the well-known criterion for instability.) As a result,
the film is not uniform any more and also changes with time.

\subsection{First iteration step}

\label{ss1} We represent the exact velocities and pressure in the form of
sums of Nusselt's solutions and the deviations from those: 
\[
w=w_N+\tilde w_0,~~~u=u_N+\tilde u_0, 
\]
\begin{equation}
v=v_N+\tilde v_0,~~~p=p_N+\tilde p_0,  \label{evp0}
\end{equation}
where tildes indicate the deviations from the Nusselt solutions. First, we
consider the $z$-momentum NS equation. We substitute the expressions (\ref
{evp0}) for velocities and pressure into Eq. (\ref{e01c}). The resulting 
(exact) equation can be written in the form
\[
\tilde w_{0xx}=R\tilde p_{0z}-\nabla ^2\tilde w_0+R\tilde w_{0t}+R\tilde u%
_0(w_N+\tilde w_0)_x 
\]
\begin{equation}
+R\tilde v_0\tilde w_{0y}+R(w_N+\tilde w_0-V)\tilde w_{0z},
\label{edew00}
\end{equation}
where $\nabla ^2=\partial ^2/\partial y^2+\partial ^2/\partial z^2$. 
In accordance with the MRD, as was discussed above, in
order to obtain a solvable ODE in $x$, we have to discard all the terms on
the right-hand-side (RHS) of (\ref{edew00}), since every one of those
contains unknown quantities. This yields a simplified equation for the
velocity $\tilde w_0$, whose solution we denote by $w_0$: 
\begin{equation}
w_{0xx}=0.  \label{edew01}
\end{equation}
We call $\tilde w_1$ the error in the approximation of $\tilde w_0$ by $w_0$%
: 
\begin{equation}
\tilde w_0=w_0+\tilde w_1.  \label{etw0}
\end{equation}

It is clear [see (\ref{eh0}) and (\ref{evc0})] that the characteristic $x$%
-lengthscale $X$ is of $O_M$ of $1$ ($X\sim 1$), i.e. (since $%
\partial /\partial x\sim 1/X$), 
\begin{equation}
\frac{\partial}{\partial x}\sim 1.  \label{edex1}
\end{equation}
We denote by $Y$ and $Z$ the characteristic lengthscales in the spanwise and
streamwise directions respectively, so that 
\begin{equation}
\frac{\partial}{\partial y}\sim \frac 1Y  \label{edey1}
\end{equation}
and 
\begin{equation}
\frac{\partial}{\partial z}\sim \frac 1Z.  \label{edez1}
\end{equation}
Similarly, denoting by $T$ the characteristic timescale, we have 
\begin{equation}
\frac{\partial}{\partial t}\sim \frac 1T.  \label{edet1}
\end{equation}
Also, clearly, $\partial ^2/\partial x^2\sim 1$, $\partial ^2/\partial
z^2\sim 1/Z^2$, $\partial /\partial y^2\sim 1/Y^2$, etc. 

The characteristic
magnitude of each of the neglected terms on the RHS of Eq. (\ref{edew00}%
)---estimated by replacing $\tilde w_0$ with $w_0$, etc.---should be smaller
than that of the term on the LHS, $w_{0xx}$: $w_0/Z^2\ll w_0/X^2$ and $%
w_0/Y^2\ll w_0/X^2$. Hence, we have the following restrictions on the
(instantaneous) lengthscales for the validity of our theory 
\begin{equation}
\frac 1{Y^2}\ll 1  \label{evc1y}
\end{equation}
and 
\begin{equation}
\frac 1{Z^2}\ll 1.  \label{evc1z}
\end{equation}
Thus, as a consequence of our derivability requirement, we have obtained the
longwave conditions---which are rather {\it postulated} in the conventional
``lubrication'' or ``long-wave'' derivations.

Similarly, using (\ref{edez1}) and (\ref{edet1}), the conditions that the
advective inertial term containing ${w}_0$, $RV{w}_{0z}$, and the
time-derivative $R{w}_{0t}$ be smaller than ${w}_{0xx}$ lead to the
following requirements on the parameters: 
\begin{equation}
\frac RZ\ll 1  \label{evc2}
\end{equation}
and 
\begin{equation}
\frac RT\ll 1.  \label{evc3}
\end{equation}
Henceforth, we confine our consideration to the film configurations which
satisfy the conditions (\ref{evc1y})-(\ref{evc3}). Note that since the local
parameters $A$, $Z$, $Y$, and $T$ can change with time, they may cease to
satisfy the validity conditions 
[such as (\ref{evc0}) and (\ref{evc1y})-(\ref{evc3})]
at a certain stage of the evolution. In such cases, clearly, the
single-EE description would be valid only for a limited time that these (and
some other, additional conditions appearing in Appendix \ref{aa1}) still
hold. Later, we will determine domains in the space of global parameters for
which such a violation of VCs never happens---so that the single-EE
approximation is valid globally rather than merely locally in time.

The boundary condition on $\tilde w_0$, at $x=h$, given by the
tangential-stress balance equation (\ref{e04b}), is written in the form 
\[
\tilde w_{0x}=-w_{Nx}-\tilde u_{0z}+[2(\tilde w_{0z}-\tilde u_{0x})\eta _z+(%
\tilde v_{0z}+\tilde w_{0y})\eta _y 
\]
\begin{equation}
+(\tilde u_{0y}+\tilde v_{0x})\eta _y\eta _z](1-\eta _y^2)^{-1}~~~(x=h).
\label{ebcw00}
\end{equation}
According to the MRD, we have to drop all the terms
containing the unknown quantities [those denoted by letters with tilde] on
the RHS. Also, using the smallness of the surface deviation (\ref{evc0}), we
will everywhere transfer the boundary conditions from the true boundary $x=h$
to a convenient ``boundary'' $x=1 $ by expanding all quantities in Taylor
series around $x=1$ (cf. Ref. \cn{h91}), such as 
\begin{equation}
\tilde{w}_{0x}(x=h)$= $\tilde{w}_{0x}(x=1)+ \tilde{w}_{0xx}(x=1)\eta+\cdots.
\label{ebcte}
\end{equation}
Noting that $w_N(x=1)=0$, we have the simplified boundary condition (for the 
{\it approximation} $w_0$ of the exact quantity $\tilde w_0$): 
\begin{equation}
w_{0x}=-w_{Nxx}\eta =2\eta ~~~~~~~(x=1).  \label{ebcw02}
\end{equation}
Finally, the no-slip condition requires 
\begin{equation}
w_0=0~~~~~~~(x=0).  \label{ensw0}
\end{equation}
The solution of Eq. (\ref{edew01}) with the boundary conditions (\ref{ebcw02}%
) and (\ref{ensw0}) is clearly 
\begin{equation}
w_0=2\eta x.  \label{ew0}
\end{equation}
Note that from (\ref{ewn}) and (\ref{ew0}) 
\begin{equation}
w_{Nx}+w_{0x}=0 ~~~ (x=h),  \label{ewn0}
\end{equation}
which we will use below. We also observe that 
\begin{equation}
w_0\sim A~(\ll w_N\sim 1).  \label{eomw0}
\end{equation}

Next, consider the incompressibility equation [see Eq. (\ref{e02})] in the
form 
\begin{equation}
\tilde u_{0x}=-\tilde v_{0y}-w_{0z}-\tilde w_{1z}.  \label{edeu00}
\end{equation}
(Note that $w_N$ does not make any contribution to this equation, since $w_N$
does not depend on $z$.) Dropping the terms with unknowns $\tilde v_0$ and $%
\tilde w_1$ on the RHS, we obtain an equation for the approximation $u_0$: 
\begin{equation}
u_{0x}=-w_{0z}=-2x\eta _z.  \label{edeu01}
\end{equation}
The no-slip BC (\ref{e03}) requires 
\begin{equation}
u_0=0~~~~~~~(x=0),  \label{ensu0}
\end{equation}
and one readily obtains the solution 
\begin{equation}
u_0=-x^2\eta _z.  \label{eu0}
\end{equation}
We note that 
\begin{equation}
u_0\sim \frac AZ\ll 1.  \label{eomu0}
\end{equation}
We denote the deviation of the exact solution $\tilde u_0$ from the
approximation $u_0$ by $\tilde u_1$, so that 
\begin{equation}
\tilde u_0=u_0+\tilde u_1.  \label{etu0}
\end{equation}

The ($x$-momentum) NS equation for the 
deviation of pressure from the Nusselt solution $p_N$
is 
\[
\tilde p_{0x}=\frac 1R (u_{0xx}+u_{0yy}+u_{0zz})
+\frac 1R (\tilde u_{1xx}+\tilde u_{1yy}+\tilde u_{1zz}) 
\]
\[
-(u_{0}+\tilde u_{1})_t-(u_0+\tilde u_1)(u_0+\tilde u_1)_x 
-\tilde v_0(u_0+\tilde u_1)_y 
\]
\begin{equation}
-(w_N+w_0+\tilde w_1-V)(u_0+\tilde u_1)_z  \label{edep00}
\end{equation}
[where we have used Eq. (\ref{edepn})].
In this equation, we have to drop the terms containing the unknowns $\tilde v%
_0$, $\tilde w_1$, or $\tilde u_1$. Next, it is easy to show that the $%
u_{0xx}$ term [which is $\sim u_0$ due to (\ref{edex1})] on the RHS is much
larger than the other terms. Namely, $u_{0yy}\sim u_0/Y^2\ll u_0\sim u_{0xx}$
due to (\ref{evc1y}) and $u_{0zz}\sim u_0/Z^2$ $\ll $ $u_0\sim u_{0xx}$ due
to (\ref{evc1z}). Similarly, $Ru_{0t}\sim (R/T)u_0$ $\ll $ $u_0\sim u_{0xx}$
by making use of (\ref{evc3}) and $R(w_N+w_0-V)u_{0z}\sim Ru_{0z}\sim
(R/Z)u_0$ $\ll $ $u_0\sim u_{0xx}$ due to (\ref{evc2}). Finally, $%
u_0u_{0x}\ll u_{0x}\sim u_{0xx}$, since $u_0\ll 1$ [see Eq. (\ref{eomu0})].
Thus the simplified equation (for the {\it approximation} $p_0$) is 
\begin{equation}
p_{0x}=\frac 1R u_{0xx}=-\frac 2R \eta _z.  \label{edep01}
\end{equation}
The BC at $x=h$ [see the normal-stress balance condition, Eq. (\ref{e05})]
is 
\widetext
\[
\tilde p_0=-p_N+\frac 2R\left \{
[(u_0+\tilde u_1)_x+\tilde v_{0y}\eta _y^2+(w_0+%
\tilde w_1)_z\eta _z^2 
-[(u_0+\tilde u_1)_y+\tilde v_{0x}]\eta _y 
+[\tilde v_{0z}+(w_0+\tilde w_1)_y]\eta _y\eta _z
\right .
\]
\[
\left .
-[(u_0+\tilde u_1)_z+\tilde 
w_{1x}]\eta _z\right \} 
(1+\eta _y^2+\eta _z^2)^{-1} 
-\sigma \left[ \eta _{yy}\left( 1+\eta _z^2\right) +\eta _{zz}\left( 1+\eta
_y^2\right)  
-2\eta _y\eta _z\eta _{yz}\right] 
\]
\begin{equation}
\times (1+\eta _y^2+\eta
_z^2)^{-3/2} ~~~ (x=h)  \label{ebcp02}
\end{equation}
\narrowtext
\noindent [where we have used (\ref{ewn0}) to 
eliminate $w_{Nx}$ and $w_{0x}$]. We
have to drop those terms containing $\tilde v_0$, $\tilde w_1$, or
$\tilde u%
_1$. In addition, a number of known terms are estimated to be smaller than
the term with $u_{0x}$. Namely, by using the estimates of $w_0$ and $u_0$, (%
\ref{eomw0}) and (\ref{eomu0}), $w_{0y}\eta _y\eta _z\sim (A/Z)(A/Y)^2\ll $ $%
A/Z\sim u_0\sim u_{0x}$. Similarly, $w_{0z}\eta _z^2\sim (A/Z)(A^2/Z^2)$ $%
\ll $ $u_{0x}$. Also, $u_{0y}\eta _y\sim u_0A/Y^2\ll u_{0x}$ and $%
u_{0z}\eta _z\sim u_0A/Z^2$ $\ll $ $u_{0x}$. Finally, one can see that
each of the other known nonlinear terms can be neglected since $\eta
_z^2\sim A^2/Z^2\ll 1$, $\eta _y^2\sim A^2/Y^2\ll 1$, and $\eta _y\eta
_z\eta _{yz}\sim A^3/(Y^2Z^2)\ll 1$. We also note that the entire Taylor
expansion for $p_N(x=1+\eta )$ consists of just one term, 
\begin{equation}
p_N(h)=p_{Nx}(1)\eta .  \label{etepn}
\end{equation}
With the known expressions for $p_N$, $w_0$, and $u_0$ (see above), and
using Taylor series about $x=1$, truncated to the first nonzero term, for
all the quantities, one obtains the BC 
\[
p_0=-p_{Nx}\eta +\frac 2R u_{0x}-\sigma \nabla ^2\eta 
\]
\begin{equation}
=\frac 2R(\cot \theta )\eta-\frac 4R \eta _z
-\sigma \nabla ^2\eta ~~~~~~(x=1).
\label{ebcp04}
\end{equation}
The solution of Eq. (\ref{edep01}) with the
boundary condition (\ref{ebcp04}) is 
\begin{equation}
Rp_0=2(\cot \theta )\eta -2\eta _z(1+x)-2W\nabla ^2\eta .  \label{ep0}
\end{equation}
We observe that 
\begin{equation}
Rp_0\sim \mbox{max }\left( \cot \theta ,~\frac 1Z,~\frac{W}{Z^2},
~\frac{W}{Y^2}\right) A.
\label{eomp0}
\end{equation}
As usual, we call $\tilde p_1$ the difference between $\tilde p_0$ and its
approximation $p_0$: 
\begin{equation}
\tilde p_0=p_0+\tilde p_1.  \label{etp0}
\end{equation}

Finally, consider the $y$-NS equation, 
\[
\tilde v_{0xx}=R(p_0+\tilde p_1)_y-\nabla ^2\tilde v_0+R\tilde v_{0t}+R(u_0+%
\tilde u_1)\tilde v_{0x} 
\]
\begin{equation}
+R\tilde v_0\tilde v_{0y}+R(w_N+w_0+\tilde w_1-V)\tilde v_{0z}.
\label{edev00}
\end{equation}
Dropping the unknown terms on the RHS, we obtain the simplified equation: 
\[
v_{0xx}=Rp_{0y} 
\]
\begin{equation}
=2(\cot \theta )\eta _y-2(1+x)\eta _{yz}-2W\nabla ^2\eta _y,  \label{edev01}
\end{equation}
where we have used the expression (\ref{ep0}) for $p_0$ in terms of $\eta $
to get the RHS in the explicit form.

The BC on $\tilde v_0$, at $x=h$, is [see Eq. (\ref{e04a})] 
\[
\tilde v_{0x}=-(u_0+\tilde u_1)_y+\{2[\tilde v_{0y} 
\]
\[
-({u}_0+\tilde u_1)_x]\eta _y
+[(u_0+\tilde u_1)_z+\tilde w_{1x}]\eta _y\eta
_z 
\]
\begin{equation}
+[\tilde v_{0z}+(w_0+\tilde w_1)_y]\eta _z\}\left( 1-\eta _y^2\right)
^{-1}~~~(x=h)  \label{ebcv00}
\end{equation}
where we have again used (\ref{ewn0}) to eliminate $w_{Nx}+w_{0x}$. Omitting
the terms with tildes and performing the by
now familiar estimates of the known
terms on the RHS, we see that the term $u_{0y}$ is larger than any other
term, so that the simplified BC at $x=1$ is 
\begin{equation}
v_{0x}=-u_{0y}=\eta _{zy}~~~~~~~(x=1),  \label{ebcv01}
\end{equation}
where the expression (\ref{eu0}) for $u_0$ has been used. Finally, the
no-slip condition 
\begin{equation}
v_0=0~~~~~~~(x=0)  \label{ensv0}
\end{equation}
is to be satisfied.
The solution of Eq. (\ref{edev01}) with the 
BCs (\ref{ebcv01}) and (\ref
{ensv0}) ---as can be readily checked by direct substitution---is 
\[
v_0=\left( \cot \theta \eta _y-W\nabla ^2\eta _y\right) (x^2-2x) 
\]
\begin{equation}
-\eta _{zy}\left (\frac{x^3}{3}+x^2-4x\right )  \label{ev0}
\end{equation}
This yields the following estimate for $v_0$: 
\begin{equation}
v_0\sim \mbox{max }\left[ \frac{\cot\theta}{Y},~
\frac{W}{Y^3}, ~\frac{W}{YZ^2},~ \frac{1}{YZ}\right] A.
\label{eomv0}
\end{equation}

As usual, 
\begin{equation}
\tilde v_0=v_0+\tilde v_1.  \label{evt0}
\end{equation}

Substituting (\ref{evt0}) into (\ref{edeu00}), we observe that in obtaining
the simplified equation (\ref{edeu01}) for $u_{0x}$, we have in fact
discarded $v_{0y}$. This requires 
\begin{equation}
v_{0y}\sim \frac{v_0}{Y}\ll u_{0x}\sim u_0  \label{evu0}
\end{equation}
(which will be used below). By using here the estimates (\ref{eomu0}) for $%
u_0$ and (\ref{eomv0}) for $v_0$, we arrive at the VCs $\mbox{max}
~ [Z\cot
\theta /Y^2,~WZ/Y^4,~W/(Y^2Z)]\ll 1$. These are a subset of the complete set of
(instantaneous) VCs obtained in Appendix \ref{aa1} [see Eq. (\ref{evct})]. It is
straightforward to verify that each discarded term in every step of our
procedure is small as a consequence of those VCs. 

The kinematic condition [Eq. (\ref{e08})] at $x=h$ becomes 
\[
\eta _t+(v_0+\tilde v_1)\eta _y+(w_N+w_0+\tilde w_1-V)\eta _z 
\]
\begin{equation}
=u_0+\tilde u_1.  \label{ekc0}
\end{equation}
Dropping the unknown terms with tilde and the smaller terms $v_0\eta _y\sim
(v_0/Y)\eta \ll u_0$ [see Eq. (\ref{evu0})] and $w_0\eta _z\sim A^2/Z\ll u_0$%
, we have 
\begin{equation}
\eta _{t_0}+(w_N-V)\eta _z=u_0,  \label{ekc00}
\end{equation}
where we have introduced the ``fast'' time $t_0$, such that $\partial
_t=\partial _{t_0}+\partial _{t_1}\approx \partial _{t_0}$. 
Using the Taylor
expansions for $w_N$ and $u_0$, we obtain, at $x=1$, 
\begin{equation}
\eta _{t_0}+(2-V)\eta _z=0.  \label{ekc01}
\end{equation}
Choosing $V=2$, we can eliminate the fast-time undulations (which are
clearly
due to the uniform translation of the wave, with no change in its shape):
\begin{equation}
\eta _{t_0}=0.  \label{et01}
\end{equation}
Thus, the leading approximation determines the velocity of a reference frame
in which film thickness does not change on the fast time scale. However,
it will change with the slower time $t_1$. In order to obtain this
slower-time evolution of the film thickness, one needs to consider the next
approximation for the velocities and pressure. From now on, we fix 
\begin{equation}
V=2.  \label{ev}
\end{equation}
Then, in view of (\ref{et01}), $\partial _t=\partial _{t_1}$.
\subsection{Second iteration}
\label{ss2} 
We now proceed to consider
the ``corrections'' $\tilde w_1$, $%
\tilde u_1$, $\tilde p_1$, and $\tilde v_1$ 
for the velocities and pressure.
By substituting 
\[
\tilde w_0=w_0+\tilde w_1, ~~~ \tilde u_0=u_0+\tilde u_1,
\]
\begin{equation}
\tilde v_0=v_0+\tilde v_1, ~~~ \tilde p_0=p_0+\tilde p_1 
\label{evp1}
\end{equation}
into the $z$-NS
equation (\ref{edew00}),
and taking into account $w_{0xx}=0$ (\ref{edew01}),
we have the exact equation 
\[
\tilde w_{1xx}=-w_{0xx}+Rp_{0z}-\nabla ^2w_0+Rw_{0t} 
\]
\[
+Ru_0(w_N+w_0)_x+Rv_0w_{0y}+R(w_N+w_0-2)w_{0z} 
\]
\begin{equation}
+%
\mbox{[terms containing $\tilde{w}_1$, 
$\tilde{u}_1$, $\tilde{v}_1$, or $\tilde{p}_1$]}.  \label{edew10}
\end{equation}
Performing our standard simplification procedure, i.e. discarding the
unknown terms (containing tilde) and small terms, and also
taking into account that $%
w_0\ll w_N$ [see Eq. (\ref{eomw0})] and $v_{0y}\ll u_0$ [see Eq. (\ref{evu0}%
)], the simplified equation for the approximation $w_1$ is 
\[
w_{1xx}=Rp_{0z}-\nabla ^2w_0+Ru_0w_{Nx}+Rw_Nw_{0z} 
\]
\[
-2Rw_{0z}+Rw_{0t} 
=2(\cot \theta \eta _z-W\nabla ^2\eta _z)-2(x+1)\eta _{zz}
\]
\begin{equation}
-2x\nabla ^2\eta 
+(2x^2-4x)R\eta _z+2xR\eta _t  \label{edew11}
\end{equation}
where we have used the known expressions (in terms of surface deviation $%
\eta $; see the preceding section) 
for the first-iteration approximations $p_0$, $w_0$, and $u_0$.
(For analogous equations of the general, $n$th iteration step,
see Appendix \ref{ac1}.)
 
The BC on $\tilde w_1$, at $x=h$, comes from (\ref{e04b}): 
\[
\tilde w_{1x}=-u_{0z}+2w_{0z}\eta _z-2u_{0x}\eta _z+w_{0y}\eta _y+v_{0z}\eta
_y 
\]
\[
+(u_{0y}+v_{0x})\eta _y\eta _z 
\]
\begin{equation}
+\mbox{[terms containing $\tilde{w}_1$, 
$\tilde{u}_1$, $\tilde{v}_1$, or $\tilde{p}_1$]} ~~~~~ (x=h)  \label{ebcw11}
\end{equation}
where we have taken into account (\ref{ewn0}). Continuing to use the
simplification procedure established in the previous section, we arrive at
the boundary condition 
\begin{equation}
w_{1x}=-u_{0z}=\eta _{zz}~~~~~~~(x=1).  \label{ebcw12}
\end{equation}
All other terms on the RHS of Eq. (\ref{ebcw11}) are readily estimated in
our usual way to be smaller than $u_{0z}$ [Eq. (\ref{evu0}) is useful in
estimating terms containing $v$]. The no-slip condition is 
\begin{equation}
w_1=0~~~~~~~(x=0).  \label{ensw1}
\end{equation}
The solution of the problem (\ref{edew11}), (\ref{ebcw12}), and (\ref{ensw1}%
) is 
\[
w_1=\left( \cot \theta \eta _z-W\nabla ^2\eta _z\right) (x^2-2x) 
\]
\[
+\left(\frac{x^4}{6}-\frac 23 x^3+\frac 43 x\right) R\eta _z
+\left( 5x-\frac 23 x^3-x^2\right) \eta
_{zz} 
\]
\begin{equation}
+\left( x-\frac {x^3}{3}\right) \eta _{yy}
+\left (\frac{x^3}{3}-x\right )R\eta _t,  \label{ew1}
\end{equation}
which can be verified by direct substitution into the problem equations.
Note that all the terms of $w_1$ are estimated to be quadratic in the
local parameters (\ref{evct}); we will see that, in general (see Appendix
\ref{ac1}), $w_n$ is of the power $(n+1)$, and similarly for 
$u_n$, $v_n$, and $p_n$.

Taking into account $u_{0x}=-w_{0z}$ [see Eq. (\ref{edeu01})], the
incompressibility condition yields the equation for $\tilde u_1$, 
\begin{equation}
\tilde u_{1x}=-w_{1z}-\tilde w_{2z}-v_{0y}-\tilde v_{1y},  \label{edeu10}
\end{equation}
where we have expressed $\tilde w_1$ as $\tilde w_1=w_1+\tilde w_2$.
Dropping the unknown terms, 
we obtain an equation for the approximation $u_1$%
: 
\[
u_{1x}=-w_{1z}-v_{0y} 
\]
\[
=-(x^2-2x)(\cot \theta \nabla ^2\eta -W\nabla ^4\eta ) 
-\left (\frac{x^4}{6}-\frac 23 x^3+\frac 43 x\right )R\eta _{zz} 
\]
\begin{equation}
-\left (5x-\frac 23 x^3-x^2\right )\nabla ^2\eta _z
-\left (\frac{x^3}{3}-x\right )R\eta _{tz}  \label{edeu11}
\end{equation}
with the BC 
\begin{equation}
u_1=0~~~~~~~(x=0).  \label{ensu1}
\end{equation}
It is easy to verify that 
\[
{u}_1=\left (\frac{x^3}{3}-x^2\right )
\left[ W\nabla ^4\eta -\cot \theta \nabla ^2\eta \right] 
\]
\[
-\left( \frac{x^5}{30}-\frac{x^4}{6}+\frac 23 x^2\right) R\eta _{zz} 
+\left( \frac{x^4}{6}+\frac{x^3}{3}-\frac 52 x^2
\right) \nabla ^2\eta _z 
\]
\begin{equation}
-\left (\frac{x^4}{12}-\frac{x^2}{2}\right )R\eta _{tz}  \label{eu1}
\end{equation}
is the solution. We note that the complete set of VCs (\ref{evct}) obtained
in Appendix \ref{aa1} guarantees that all the terms discarded in obtaining
solvable ODEs for $w_1$ and $u_1$ are small in comparison with (the biggest
of) those terms that are retained.

At this point, we could proceed to solve the $x$-NS and $y$-NS equations for
the pressure and velocity corrections $\tilde p_1$ and $\tilde v_{1\text{ }}$%
respectively. However, these corrections are not needed for obtaining the
second-iteration EE. (We will only need the pressure and velocity
corrections for {\it later} iteration stages, and we calculate those corrections
in Appendix \ref{ab1}.)

\section{The dispersive-dissipative evolution equation}

\label{s4} The (exact)
kinematic condition (\ref{e08}) at $x=h$ can be written in
the form 
\[
\eta _t+(v_0+\tilde v_1)\eta _y+(w_N+w_0+w_1+\tilde w_2-2)\eta _z 
\]
\begin{equation}
=u_0+u_1+\tilde u_2 ~~~~~ (x=1+\eta )  \label{ekc10}
\end{equation}
where we have used $V=2$ (\ref{ev}). Dropping the terms containing unknown
velocities (those with tilde) and using the Taylor series to relate the
velocity components at $x=h$ to those at $x=1$, we have 
\begin{equation}
\eta _t+(w_{Nx}\eta +w_0)\eta _z=u_{0x}\eta +u_1~~~(x=1).  \label{ekc1}
\end{equation}
In (\ref{ekc1}), we have dropped the terms $v_0\eta _y$ and $w_1\eta _z$ as
they are smaller than $u_1$ [see Eq. (\ref{edeu11})]. Also, recall that $%
w_{Nx}(x=1)=0$ [see Eq. (\ref{ewn})]. 
Using the expressions (\ref{ew0}), (\ref
{eu0}), and (\ref{eu1}) for $w_0$, $u_0$, and $u_1$, we obtain 
\[
\left [ \eta -\frac{5}{12}R\eta _z\right ]_t
+4\eta \eta _z+\frac 23\delta \eta _{zz}-%
\frac 23\cot \theta \eta _{yy} 
\]
\begin{equation}
+\frac 23W\nabla ^4\eta +2\nabla ^2\eta _z=0  \label{eee00}
\end{equation}
where by definition 
\begin{equation}
\delta \equiv \frac 45 R-\cot \theta .  \label{del=}
\end{equation}
However, the term $\propto$
$R\eta _{tz}$ $\sim ~(R/Z)\eta _t$ $\ll \eta _t$
since $R/Z\ll 1$ [see (\ref{evct})]. Dropping this small term, we have 
\[
\eta _t+4\eta \eta _z+\frac 23\delta \eta _{zz}-\frac 23
\cot \theta \eta_{yy}
\]
\begin{equation} 
+\frac 23W\nabla ^4\eta +2\nabla ^2\eta _z=0.  \label{eee10} 
\end{equation} 
 
Simple linear-stability analysis can reveal the dynamical role of some terms
here. Assuming an infinitesimally small disturbance in the form of a normal
mode, $\eta \propto \exp (st-iwt)\exp i(jy+kz)$, it readily follows from the
linearized version of (\ref{eee10}) that 
\begin{equation}
s=\frac 23[\delta k^2-(\cot \theta )j^2-W(k^4+2j^2k^2+j^4)]  \label{elsa}
\end{equation}
and
\begin{equation}
\omega =-2k(k^2+j^2).  \label{eloa}
\end{equation}
Here $s$ is the growth (or decay) rate for the disturbance and so the third,
fourth, and fifth terms in Eq. (\ref{eee10}) which give rise to growth (or
decay) are {\it dissipative}
[considering the destabilizing term (the one with $%
\delta $) as a {\it negative} dissipation]. 
In contrast, the last term in (\ref
{eee10}) only makes a contribution to the (real) frequency $\omega $, rather
than to the growth rate $s$, i.e. it does not lead to growth or decay of
disturbances. Thus, this (third-derivative) term is 
{\it dispersive}. 

Clearly, for
instability to develop (i.e. for $s>0$), we need 
\begin{equation}
\delta >0,  \label{del>}
\end{equation}
a condition we assume fulfilled from now on. This yields the so-called
critical value $R_c$ of the Reynolds number, $R_c=(4/5)\cot \theta ,$ at which
the instability sets in.

One can see from the above derivation of the 
dispersive-dissipative EE (\ref{eee10}) that the
destabilizing (third) term originates from the inertia terms of the NS
equations. The (stabilizing) fourth and fifth terms are due respectively to
hydrostatic and capillary (i.e. surface-tension) parts of the pressure.
Finally, the last, odd-derivative term is due to the viscous part of the
pressure. Such a purely {\it dispersive }term also appeared in the EE
obtained by Topper and Kawahara \cite{tk78} for an almost vertical plane:
they used the small angle of the plane with the vertical as their (single)
perturbation parameter (see also the discussion in the Introduction of the
present paper). Our derivation shows that assumption to be unnecessary. In
particular, for the vertical film $\cot \theta =0$, and Eq. (\ref{eee10})
becomes 
\begin{equation}
\eta _t+4\eta \eta _z+\frac 8{15}R\eta _{zz} 
+\frac 23W\nabla ^4\eta +2\nabla ^2\eta _z=0.  \label{eee10v}
\end{equation}
Although an equation of this structure (but with arbitrary coefficients) was
postulated as a model equation in Ref. \cn{ti89}, it cannot be obtained from
the derivation of Topper {et al.}\cite{tk78}:
Since their small parameter is proportional to 
$\cot \theta $, it becomes zero for the vertical case, and the Reynolds
number (also proportional to the small parameter in that SP derivation)
vanishes---which, clearly, cannot correspond to any flow at all.

The (infinite-dimensional) dynamical system governed by the {\it dissipative}
equation (\ref{eee10}) essentially 
forgets initial conditions as it evolves
towards an attractor. There may be 
fluctuations on the attractor, but there is no
systematic change in time. Clearly then the amplitude-decreasing,
stabilizing term must balance the destabilizing one (the latter
tends to increase
the deviation amplitude). So the two dissipative terms are 
necessarily of the same order
of magnitude on the attractor.

As to the magnitude of the dispersive term relative to that of
the dissipative terms, there can occur, depending on location in the
parameter space, each of the following three possibilities: (1)
these terms are of the same $O_M$; (ii) the dispersive term is small (and
then the amplitude is determined by the balance of the nonlinear term with
the dissipative terms); and (iii) the dissipative terms are small (and the
nonlinear term balances the dispersive one). In the first case, continuing
the iteration process would lead to small corrections to the terms which
cannot significantly change the evolution. In the second case, the small
dispersive term can be omitted with a negligible effect, so the corrections
would be again immaterial. 

But in the third case, when dissipation is small,
the situation is very different. Discarding the dissipative terms leads to a
2D Korteweg-deVries (KdV) equation which was simulated numerically in 
Ref. \cn{it90}. The KdV equation is purely 
dispersive and never forgets the initial
conditions. It has a one-parameter family of axisymmetric solutions which
are traveling solitons, similar to the well-known 1D KdV case. Depending on
the initial state, there may be solitons of different lengthscales (and
therefore moving with different speeds) in the final state. However, if the
small dissipative terms are present, they will slowly change the initial
soliton of an arbitrary lengthscale. It will evolve along the soliton family
until the lengthscale is attained which provides for the balance between the
two dissipative terms (this effect was first studied for the 1D case in
Ref. \cn{k83}; see also Refs. \cn{os69,os70,ad90}). 
Therefore, the dissipative terms, even when {\it small}, are
important: they determine the lengthscale of the solution. 

However, only the
{\it largest-magnitude} terms are guaranteed to 
be correct in the above beginning-iteration
derivation; as for the smaller terms, further iterations might yield
significant corrections to them. We consider this question (of higher
iterations and corrections to small dissipative terms) in the next section.
It turns out (perhaps, surprisingly) that such corrections can be important
{\it locally} under some parametric conditions, 
but that no (single) corrected EE can
approximate the evolution for all time. 
Equation (\ref{eee10}) is thus the most
general of those EEs that can be valid {\it globally}
in time---under appropriate parametric
restrictions, which can be completely determined only with the analysis of
higher iterations of the NS problem, as is done in the next section. For the
rest of this section, we continue the consideration of the GEE (\ref
{eee10}).

From the condition $\delta >0$, Eq. (\ref{del>}), it follows that 
\begin{equation}
R>\frac 54\cot \theta >\cot \theta .  \label{ercot}
\end{equation}
From Eqs. (\ref{eee10}) and (\ref{elsa}), it is clear that in order for
instability to develop, we need $\delta \eta _{zz}>(\cot \theta)\eta _{yy}$.
Using the $y$ and $z$ lengthscales, this yields, $Y^2/Z^2>(\cot \theta
)/\delta $. Noting that either $\delta \ll R\sim R_c\sim \cot \theta $ or $%
\delta \sim R$, we see that $Y\gg Z$ or $Y\sim Z$, except perhaps for $R\gg
\cot \theta $. For simplicity, we assume that $Z\leq Y$, which seems to be
the case in all experiments we know about. Then 
\begin{equation}
L\equiv \mbox{min}(Z,~Y)=Z.  \label{el}
\end{equation}
With this and the condition (\ref{ercot}), the VCs (\ref
{evctl}) reduce to 
\begin{equation}
\mbox{max}\left[ A, ~\frac{1}{L^2}, ~\frac RL,
~\frac{W}{L^3}\right] \ll 1.  \label{etvcn}
\end{equation}
These are conditions of {\it instantaneous}
validity; they involve the local (i.e. instantaneous) parameters $%
A(t)$ and $L(t)$. 

As was discussed above, due to the dissipativeness of the EE (\ref{eee10}),
the system evolves towards an attractor, and in the asymptotic limit of
large times we have 
$A(t)=const\equiv A_a$ and $L(t)=const\equiv L_a$. Since (similar
to Refs. \cn{bf83,fb87}) the 
destabilizing inertia term should be of the same $%
O_M$ as the stabilizing, capillary one, i.e. $\delta \eta _{zz}(\sim \delta
A/L_a^2)\sim W\nabla ^4\eta (\sim WA/L_a^4)$, the (dimensionless)
characteristic lengthscale {\it at large times} $L_a$ can be taken to be 
\begin{equation}
L_a=\left( \frac {W}{\delta} \right) ^{1/2}.  \label{els}
\end{equation}
Similarly, the asymptotic magnitude of the characteristic amplitude $A_a$ is
determined by the balance between the nonlinear ``advective'' term and
either the dispersive term or the capillary one (whichever is larger): $A_a=%
\mbox{max}(W/L_a^3,~1/L_a^2)$. Using these asymptotic values of parameters,
the conditions (\ref{etvcn}) can be written as $\mbox{max}%
(W/L_a^3,~R/L_a,~1/L_a^2)\ll 1$. Noting that in view of Eq. (\ref{els}), $%
W/L_a^3=\delta /L_a$ and [see Eq. (\ref{del=})] $R=(5/4)(\delta +\cot \theta
)>\delta $, so that $W/L_a^3<R/L_a$, we can simplify the VC to $\mbox{max}%
\left( R/L_a,~L_a^{-2}\right) \ll 1$; in terms of the basic parameters, 
\begin{equation}
\alpha \equiv \frac{1}{L_a^2}=\frac{\delta}{W}\ll 1,
~~\beta \equiv \frac{R}{L_a}=R\left (\frac{\delta}
{W}\right )^{1/2}\ll 1.  \label{evcg0}
\end{equation}
In the next section, it is shown that we also need 
\begin{equation}
\gamma \equiv \frac{\max(R,R^3)}{W}\ll 1  \label{evci}
\end{equation}
(otherwise, the dissipative terms contributed by higher iterations can
become significant, and the evolution cannot be all-time-describable by a
single EE). All three parameters, $\alpha $, $\beta $, and $\gamma $, are
small if (recall that $\delta <R$)

\begin{equation}
\alpha _R\equiv \frac RW=\frac{\bar \rho \bar g^2\bar h_0^5\sin ^2\theta}
{2\bar 
\sigma \bar \nu ^2}\ll 1  \label{evcg1}
\end{equation}
and 
\begin{equation}
\beta _R\equiv \frac{R^{3/2}}{W^{1/2}}
=\left (\frac{\bar \rho \bar h_0^{11}}{8\bar \sigma} \right )^{1/2}
\left (\frac{\bar g^2\sin ^2\theta }{\bar \nu ^3}\right )
\ll 1.  \label{evcg2}
\end{equation}
So, if we are in the domain of the space of basic parameters which satisfies
the condition $\max (\alpha _R,\beta _R)\ll 1$ [or a bit more general, but
less simple condition $\max (\alpha ,\beta ,\gamma)\ll 1$], then Eq. (\ref
{eee10}) is good {\it for all time}
[provided the initial amplitude and
lengthscale satisfy the conditions (\ref{etvcn})]. Therefore, we call such
conditions the ``global'' VCs.

We can transform the GEE (\ref{eee10}) to a ``canonical'' form---which
contains only two ``tunable'' constants---by rescaling the variables with
appropriate units: 
\[
\eta =N\widetilde{\eta },~~~z=L_a\widetilde{z}, 
\]
\begin{equation}
y=L_a\widetilde{y},~\mbox{and}~t=T_a\widetilde{t},  \label{escal}
\end{equation}
where $N=1/(2L_a^2)$ and $T_a=L_a^3/2$. Dropping the tildes in the notations
of variables, the resulting canonical EE is 
\begin{equation}
\eta _t+\eta \eta _z-\kappa \eta _{yy}+\nabla ^2\eta _z+\epsilon \left( \eta
_{zz}+\nabla ^4\eta \right) =0.  \label{eee12}
\end{equation}
The control parameters in this equation are 
\begin{equation}
\epsilon =\frac 13{\sqrt{W\delta }}
\end{equation}
and
\begin{equation}
\kappa =\frac{L_a \cot \theta}{3}=\frac 13 \sqrt{\frac{W}{\delta}}
\cot\theta
\end{equation}

For the case of $\kappa =0$ (e.g. flow down a vertical wall), Eq. (\ref
{eee12}) becomes essentially the model equation postulated and numerically
studied in Ref.
\cn{ti89}. If, in addition, $\varepsilon =\kappa =0,$ we have
the 2D KdV equation\cite{zk74}, whose 1D ($\partial _y=0$) version is the
usual KdV equation. When $\kappa =0$ but 
$\epsilon \rightarrow \infty $ (so
that, after an appropriate rescaling, the dispersive term disappears from
the equation), Eq. (\ref{eee12}) becomes the one obtained by Nepomnyashchy 
\cite{n74}, whose 1D version is the Kuramoto-Sivashinsky equation \cite
{kt76,s77}. All these equations are thus limiting cases of EE (\ref{eee12}).

Numerical simulations \cite{if95} of this equation 
have shown that its
solutions remain bounded for all time (which property always
is implicitly {\it assumed} in
global-validity considerations, and therefore must be directly
verified {\it after} the EE has been obtained). 
For example, Fig. \ref{f1} shows 
evolution of the surface deviation ``energy'' $\int \eta ^2 dy dz$ 
from an initial small-amplitude 
($\eta \sim 10^{-2}$) ``white-noise'' condition
to a statistically steady
state (for 
$\kappa=0$ and $\epsilon=1/50$).
Detailed numerical studies of Eq. (\ref{eee12})
will be presented in a sequel to this paper.

\section{Additional dissipative terms}

\label{s5} 
In this section, we examine the implications of the additional
dissipative terms arising from further iterations. We take into account
explicitly all the linear dissipative and dispersive terms with derivatives
of order four or less by going through the iterative process twice. We also
estimate the effect of dissipative terms with derivatives of order six
or more on the EE.
 
We note that for obtaining the evolution equation, we need only
the successive iterates of the normal velocity $u$. This is because the
nonlinear terms involving $w\eta _z$ and $v\eta _y$ in the
kinematic condition make smaller
contributions to the EE. 
(Indeed $u\sim w_z+v_y$, and e.g. $w_z\eta$ generates the same terms
as the $w_z$ part of $u$, but with the extra small factor $\eta$.)
We have found the next two
iterative corrections for the normal velocity, $u_2$ and $u_3$,
Eqs. (\ref{eu2})
and (\ref{eu3}) in Appendix \ref{ab1}.
[We need $u_3$ (in addition to $u_2$) because its dissipative terms
are not guaranteed to be much smaller than those of $u_2$. However, the
dissipative terms of $u_4$ {\it are} much smaller than those of $u_2$,
so we do not have to consider $u_4$.] 
 Using these in the kinematic
condition (\ref{e08}), we obtain [see Appendix \ref{ab1} for details] 
\widetext
\[
\eta _t=\left
[\frac{5}{12}R\eta _z-\frac{4}{15}R\cot \theta \nabla ^2\eta 
+\frac{295}{672}R^2\eta
_{zz}\right ]_t 
-4\eta \eta _z-\frac 23\delta \eta _{zz}
+\frac 23\cot \theta \eta _{yy}-2\nabla
^2\eta _z-\frac 23 W\nabla ^4\eta 
-\left [\frac{23}{15}R-2\cot \theta\right ]
(\eta \eta _z)_z
\]
\begin{equation}
+\frac{5}{14}R\cot \theta \nabla ^2\eta
_z 
-\frac 27 R^2\eta _{zzz}+\frac 65\cot \theta \nabla ^4\eta 
-\frac{331}{168}R\nabla ^2\eta
_{zz} 
-\frac{1241483}{8108100}R^3\eta _{zzzz} 
+\frac{477523}{2494800}R^2\cot \theta \nabla ^2\eta _{zz}.  
\label{eee23}
\end{equation}
\narrowtext
\noindent We get rid of time-derivatives in 
RHS by twice iterating this equation,
substituting (for each time-derivative in the RHS) the RHS 
(\ref{eee23}) of (\ref{eee23}) itself;
the remaining time-derivatives in the final RHS are omitted, because further
iterations would only lead to derivatives of an 
order higher than four, with
small resulting contributions. 
[In fact, in the first iteration, when
substituting into the (mixed-derivative)
terms with the second spatial derivatives, it is
enough to retain from the RHS (\ref{eee23}) only the terms with {\it no}
time-derivatives, and with space-derivatives of the
order two only. As for the
term with the first spatial derivative, $\propto \eta _{tz}$, the terms
without time-derivatives and with space-derivatives of orders two and three
are substituted into it, and also the term $(5/12)R\eta _{tz}$ itself, with
the result $(5/12)^2R^2\eta _{tzz}$. In the second iteration, it is
sufficient to only substitute into the latter term, and---since it already
contains two spatial differentiations---only the 
purely spatial second-derivative
terms should be substituted into it.] 
As a result, we obtain the following equation:
\widetext
\[
\eta _t+4\eta \eta _z+\frac 23\delta \eta _{zz}-
\frac 23\cot \theta \eta _{yy} 
+2\nabla ^2\eta _z+\frac{40}{63}R\delta \eta _{zzz}
-\frac{40}{63}R\cot \theta
\eta _{zyy} 
+\frac 23 W\nabla ^4\eta 
-\frac 23\cot \theta \nabla ^4\eta +\frac{157}{56}R\nabla
^2\eta _{zz}
\]
\begin{equation}
+\frac{8}{45}R\cot ^2\theta \nabla ^4\eta 
+\frac{1213952}{2027025}R^3\eta _{zzzz} 
-\frac{138904}{155925}R^2\cot \theta \nabla ^2\eta _{zz}
+\left (\frac {16}{5}R-2\cot\theta\right )(\eta \eta_z)_z=0.
\label{eee23n}
\end{equation}
\narrowtext
\noindent The 1D ($\partial _y=0$) limit of this 
equation coincides with the
small-amplitude limit of the EE obtained, with the same numerical
coefficients, in Ref.
\cn{n75}, but our 2D version is new. (We remark that
there are several mistakes in the presentation of steps 
leading to the final
equation, Eq. (27) in Ref. \cn{n75}. However, Eq. (27) itself appears to be
correct. The same numerical coefficients appeared in an even earlier paper
\cite{b66} in a linearized 1D context.) The (2D) terms with derivatives of
order three or less agree with Ref. \cn{r70}, and those plus the
surface-tension ($W$) term---with Ref. \cn{kl77}. 
However, note that some terms of (\ref
{eee23n}) are always negligible. For example, the two dispersive
third-derivative terms containing $R$ are clearly smaller than the
corresponding second-order dissipative terms (because of the 
instantaneous
VCs $%
R/L\ll 1$ and $\cot \theta /L\ll 1$), and therefore are negligible in all
cases.

As was mentioned above, since we are only interested in situations where
persistent nonlinear waves are present, $\delta >0$, we have either $\delta
\sim R$ or $\delta \ll R$ (but not $\delta\gg R$; see Eq. (\ref{del=}). 
When $\delta \sim R$, the additional
fourth-derivative terms in (\ref{eee10}) are each smaller than the
destabilizing second-derivative term (because of $R/L\ll 1$ and/or $1/L^2\ll
1$). If $\max (R,R^3)\ll W$ [see  Eq. (\ref{evci})], 
those terms are much smaller
than the stabilizing {\it capillary }term, and we return to the GEE (%
\ref{eee10}) with global VCs (\ref{evcg1}) 
and (\ref{evcg2}) (this holds as well
even for $\delta \ll R$). But if $W$ is not large enough,
so that Eq. (\ref{evci}) is violated and the capillary fourth-derivative term
is much smaller than (at least) one of the non-capillary fourth-derivative
members, then the destabilizing term cannot be balanced, and the EE leads to
the unlimited growth of amplitude. Thus, clearly, at such parameter-space
locations the EE (\ref{eee23n}) {\it can}
 be valid locally, i.e. for a limited
time only, but it is {\it not} globally-good.

Consider now (for the rest of this section)
the complementary case $\delta \ll R$ (so $R\simeq R_c$), and
with $W$ sufficiently small, so that the condition $\max (R,R^3)\ll W$ is
violated. We have obtained, except for numerical coefficients, all
the essential terms (which turn out to be linear; see the end of
Appendix \ref{ac1}).
Including these terms in (\ref{eee23n}), the EE 
can be written in
the general form 
\[
\eta _t+4\eta \eta _z+\frac 23\delta \eta _{zz}
-\frac {8}{15}R_c\eta _{yy}
+2\eta _{zzz}+\frac 23 W\eta_{zzzz}
\]
\[
+\frac 85 R_c(\eta \eta _z)_z
+\left [\frac{2581}{1400}R_c-\frac{32}{3378375}R_c^3
\right ]\eta _{zzzz}
\]
\begin{equation}
+\mbox{sum of terms of type }R_c^{2k+1}\eta^{(2l)}~=0.  \label{eee25}
\end{equation}
Here the superscript on $\eta $, enclosed in parentheses, refers to the
order of the spatial derivative; $l>k\geq 0$; $k=0,1,2,\cdot \cdot \cdot $; $%
l=(k+1),(k+2),\cdot \cdot \cdot $; and every ($k+l$) is odd. In (\ref{eee25}), we
have used the leading Taylor-series approximation putting $R=R_c$ (recall
also $\cot \theta \approx 4R_c/5)$. Also, $Y\gg Z$ 
(as a consequence of $\delta
\eta _{zz}\ge R_c\eta _{yy}$, which is required for instability to develop).
Furthermore, we have not included additional nonlinear or dispersive
terms in Eq. (\ref{eee25}):
the dissipative nonlinear terms can be shown (see Appendix \ref{ac1}) to
be smaller than the leading nonlinear dissipative term 
$\propto R(\eta \eta _z)_z,$
and all the dispersive terms are smaller than the linear one $\nabla
^2\eta _z$. 

Normally, the coefficients of the terms are $O_M(1).$ However,
sometimes a coefficient is $\ll 1$ because of an accidental near
cancellation of terms, as e.g. is the case for the term $\propto R_c^3\eta
_{zzzz}$ in (\ref{eee25}). Then we say the term is ``degenerate''. Comparing
the additional dissipative terms with the term $\propto R_c^3\eta _{zzzz}$
and taking into account the instantaneous
 VCs $R/L\ll 1$ and $1/L^2\ll 1$, the only
terms which may not be negligible are those of the structure $R_c^{2n-1}\eta
^{(2n)}$ $(n=2,3,\cdot \cdot \cdot )$ (even those could have been neglected
in comparison with the term $\propto R_c^3\eta _{zzzz}$ were the latter
non-degenerate). 
Hence, the EE can be simplified: 
\[
\eta _t+4\eta \eta _z+\frac 23\delta \eta _{zz}
-\frac{8}{15}R_c\eta _{yy}+2\eta
_{zzz}
\]
\[
+\frac{8}{5}R_c(\eta \eta _z)_z
+\left [\frac{2581}{1400}R_c
-\frac{32}{3378375}R_c^3\right ]\eta_{zzzz}
\]
\begin{equation}
+\sum_{n=3}^\infty c_nR_c^{2n-1}\eta ^{(2n)}=0,  \label{eee25s}
\end{equation}
where $c_n$ are numerical coefficients. Can this equation be valid for all
time? For this to be the case, the destabilizing term $(2/3)\delta \eta _{zz}
$ has to be balanced by a stabilizing one---by the term $\propto R_c\eta
_{zzzz}$ (the other, degenerate fourth-derivative term is clearly
destabilizing) or by the first non-degenerate and stabilizing
higher-derivative term of (\ref{eee25s}), whichever term is dominant.

Suppose first that the fourth-derivative term is the dominant stabilizing
one. Then Eq. (\ref{eee25s}) becomes 
\[
\eta _t+4\eta \eta _z+\frac 23\delta \eta _{zz}
-\frac{8}{15}R_c\eta _{yy}+2
\eta _{zzz}
\]
\begin{equation}
+\frac 85 R_c(\eta \eta _z)_z
+\frac{2581}{1400}R_c\eta _{zzzz}=0.  \label{eee26}
\end{equation}
The balance of the destabilizing term $\delta \eta _{zz}$ with the
stabilizing term $R\eta _{zzzz}$ yields the lengthscale $L\sim R_c/\delta $.
Comparing the dissipative term with the dispersive term, we have $R_c\eta
_{zzzz}/\eta _{zzz}\sim R_c/L\ll 1$: the dispersive term is always dominant.
Therefore, it is the dispersive term which is to balance the nonlinear one,
which yields the characteristic amplitude $\sim 1/L^2$. Using this, it is
easy to see that the nonlinear dissipative term $\sim R_c(\eta \eta _z)_z$
is exactly $O_M$ of the linear dissipative terms; therefore, the nonlinear
term plays a significant role. It is destabilizing, and numerical
simulations indicate that the solutions of (\ref{eee26}) blow up [see Fig. 
\ref{f2}, which shows the blow-up of 
energy (in contrast to Fig. \ref{f1}) 
governed by the equation rescaled to the canonical form similar
to Eq. (\ref{eee12}),
\[
\eta_t+\eta\eta_z+\eta_{zzz}-\kappa\eta_{yy} 
\]
\begin{equation}
+\varepsilon \left [ \eta_{zz}+\eta_{zzzz}+(1120/2581)(\eta\eta_z)_z
\right ]=0,
\label{eee12n}
\end{equation}
with $\kappa=0$ and $\varepsilon=1$; we have observed similar blow-up
behavior with all values of
$\varepsilon$ we tested in the range $10^{-2}$-$10^2$].
The thickness of the real film, of course,
is bounded uniformly for all time.
Hence, Eq. (\ref{eee26}) (which {\it can} 
be good for a limited time) is {\it not}
globally-valid. Physically, we believe that the growth of amplitude will be
arrested by viscosity after small lengthscales develop, which would violate
the small-$R/L$ VC for the single-EE description. So the EE (\ref{eee26})
cannot be valid for large times.

It remains to consider the hypothetical case when a higher-derivative
term $\propto $ $R_c^{2n-1}\eta ^{(2n)}$ is the dominant stabilizing one.
[In particular, this implies that $3\leq n=m$ (say), $c_n\ll 1$ for $n<m,$ $%
c_m\sim 1,$ and $c_m$ has an appropriate sign: $c_m>0$ if $m$ is even and $%
c_m<0$ if $m$ is odd.] We do not believe this actually happens: such terms
are traced back to be due to the inertia terms in the momentum NS equations,
the same inertia terms which give the destabilizing second-derivative term
of Eq. (\ref{eee25s}), and it seems unlikely that the same physical cause
can be responsible for both stabilization and destabilization. Therefore, we
believe the first nondegenerate term will turn out to be 
{\it destabilizing}. 

Based on the (linear) studies \cite{d90} of the Orr-Sommerfeld equation, we
have computed the next coefficient, which shows that, albeit stabilizing,
the sixth-derivative term is again degenerate: $%
c_3=-16173184/1718663821875\approx -0.941\times 10^{-5}$. (This is also
remarkably close to the fourth-derivative coefficient, $c_2\approx
-0.947\times 10^{-5}$. 
If further $c_n$ were all degenerate too and of the same
order of magnitude, the destabilizing fourth-derivative term would be
dominant, and hence stabilization and all-time valid EE impossible.). 

A special case to consider is when the destabilizing fourth-derivative
term (which is clearly much greater than the sixth-derivative one)
is nearly cancelled by the stabilizing 4th-derivative term 
$aR_c\eta^{(4)}$ (where $a\equiv 2581/1400$). 
As the term $c_3R_c^5\eta^{(6)}$ is stabilizing, 
one\cite{ks96} can ask whether there 
can be an all-time valid EE if $|c_3R_c^5\eta^{(6)}|
\gg |aR_c-c_2R_c^3)\eta^{(4)}|$. 
Our answer to this question is as follows. In this case, 
$aR_c\eta^{(4)}$ $\sim$ $|c_2|R_c^3\eta^{(4)}$ 
$\gg$ $|c_3|R_c^5\eta^{(6)}$. Since the dispersive 
(third-derivative) term must dominate the fourth-derivative
term $aR_c\eta^{(4)}$, it dominates even stronger the
sixth-derivative term $c_3R_c^5\eta^{(6)}$
[see the discussion following Eq. (\ref{eee26})]. 
As usual, the lengthscale is determined by a
balance between the dominant linear dissipative terms, $\delta\eta_{zz}$
$\sim$ $|c_3|R_c^5\eta^{(6)}$ $\ll$ $aR_c\eta^{(4)}$
$\sim$ $R_c(\eta\eta_z)_z$ [see the arguments following 
Eq. (\ref{eee26})]. Thus, the destabilizing nonlinear 
term is the greatest one; there is no other term which could serve
as a counterbalance. We arrive at the conclusion that the hypothetical
EE with the sixth-derivative term cannot be valid for all time. 

We
have not attempted to determine the next 
coefficient, $c_4$, because of the
large volume of calculations which 
would be required. Based on what we have
said above, we expect the eighth-derivative term to be nondegenerate 
and destabilizing, and then no
single EE can be globally valid under the 
circumstances. It follows that the GEE (%
\ref{eee10}) {\it is} the most general one. 
The validity condition $W\gg R_c^3$
(which would be sufficient even if $c_2$ were nondegenerate) can be relaxed
a bit: it is enough to require that the capillary term dominates the
(presumably nondegenerate) eighth-derivative one, $W/L^3\gg R_c^7/L^8.$
(With $L^2\sim W/\delta $ [see Eq. (\ref{els})], we get $W^{7/2}/(\delta
^{5/2}R_c^7)\gg 1.$) The EE (\ref{eee25s})
 is valid locally only, under the instantaneous VCs
(\ref{etvcn}).
 
If, however, the first nondegenerate term $\propto $ $R_c^{2n-1}\eta ^{(2n)}$
turned out to be stabilizing, the EE (\ref{eee25s}) would have a domain of
global validity, albeit a very limited one. It is straightforward to find
the corresponding global VCs. Indeed, the balance between this term and the
destabilizing one, $R_c^{2m-1}\eta ^{(2m)}\sim \delta \eta _{zz}$, yields
the lengthscale $L$,

\begin{equation}
L^{2m-2}\sim \frac{R_c^{2m-1}}{\delta} .  \label{elsr5}
\end{equation}
Using this lengthscale, the ``modified-$R$'' VC takes the form 
\begin{equation}
\frac{R_c}{L}\sim \left 
(\frac{\delta}{R_c}\right )^{1/(2m-2)}\ll 1,  \label{em-r}
\end{equation}
and the small-slope VC becomes
\begin{equation}
\frac{1}{L^2}\sim \left [\frac{\delta}{R_c{}^{(2m-1)}}\right
]^{1/(m-1)}\ll 1.  \label{es-s}
\end{equation}
The (relaxed) conditions of dominance are $W\eta ^{(4)}\ll R_c^{2m-1}\eta
^{(2m)}$ and $R_c\eta ^{(4)}\ll R_c^{2m-1}\eta ^{(2m)}$, i.e. $W\ll
R_c^{2m-1}/L^{2m-4}$ and $R_c^2(R_c/L)^{2m-4}\gg 1$ (which with (\ref{elsr5}%
) are easy to recast in terms of the global parameters only). The ratio of
the nonlinear dissipative term, $\propto R_c(\eta \eta _z)_z$, to the
stabilizing one, $\propto R_c^{2m-1}\eta ^{(2m)}$,
is $A/(R_c/L)^{2m-2}$,
which is required to be small---otherwise, we can have the  blow-up of
solutions, similar to the case with the term  $\propto R_c\eta ^{(4)}$ being
dominant. The comparison of dissipative term,
$\propto R_c^{2m-1}\eta ^{(2m)}$,
to the dispersive term, $\propto \eta _{zzz}$,
shows that the latter can be
greater or smaller than 
the former depending upon whether $R_c^{2m-1}/L^{2m-3}%
\ll 1$ or $R_c^{2m-1}/L^{2m-3}\gg 1$.

When $R_c^{2m-1}/L^{2m-3}\ll 1$ (dispersion is large), the characteristic
amplitude, obtained by balancing the term $\eta \eta _z$ with the dispersive
term, is $1/L^2$ [hence, the small-amplitude VC, $A\ll 1$,
coincides with (\ref
{es-s})]. Otherwise, i.e. if $R_c^{2m-1}/L^{2m-3}\gg 1$, the amplitude is
determined by balancing the term $\propto \eta \eta _z$ with the dominant
stabilizing term, $A\sim (R_c/L)^{2m-1}$. Using the global-parameter
estimate of the 
lengthscale, Eq. (\ref{elsr5}), all the above conditions are
readily reduced to certain global VCs (expressed in terms of global
parameters only); we do not write them here, in view of the likely
nonreality of the imaginary case of dominance of the term $\propto
R_c^{2n-1}\eta ^{(2n)}$.

\section{Summary}

\label{s6}
We have considered flow of a liquid film down an inclined plane. We
have posed and studied the following questions: What are the least
restrictive parametric conditions for which the wavy film flow  can be
approximated for all time by a single (local)
evolution equation, and what (if any) is the
most general form of such an equation?
 
We have argued that the dissipative-dispersive 
evolution equation (\ref{eee10})
[which we derived by an iterative perturbation method of a multiparameter
type] is such a general EE. Any all-time valid EE derived by a
single-parameter technique is necessarily nothing else but
essentially the general EE in which some terms
have been omitted. Also, the domain of validity of 
such a ``partial'' EE is necessarily a
{\it subdomain} of the ``umbrella''
domain of global validity given by Eqs. (\ref{evcg1}%
) and  (\ref{evcg2}). [In particular, in 
such domains the amplitude of waves is
necessarily much smaller than the mean film thickness.] 

It is clear that any evolution equation which follows from a 
multiparametric approach (such as the iterative technique we have
employed in this paper) can be also obtained with the conventional SP
approach. However, the significant advantage of the MP derivation is
that it covers at once all possible SP derivations of EEs (the number
of which is, in principle, infinite in the SP approach, corresponding
to the different choices of the small-parameter powers for the system
parameters). Also, comparing the two derivations of even  
a {\it particular} EE, the MP derivation is justifiable for much less
restrictive domains of the parameter space.

The theory also yields the explicit approximate expressions in terms of film
thickness for the pressure and components of velocity [Eqs. (\ref{ep0}), 
(\ref
{eu0}), (\ref{ev0}), and (\ref{ew0})], 
and thus a complete description of film dynamics.

We have derived an EE (\ref{eee23n}) containing 
additional high-derivative
terms which can be essential near the threshold of instability, $R\approx
R_c,$ if the surface tension is sufficiently small. However, this EE (\ref
{eee25s}) is good for a limited time only and cannot be good for all time.
[The conditions of such a local (in time) validity are given by Eq. (\ref
{etvcn}).] Under certain parametric conditions, the (numerical) solutions of
that EE blow up due to a nonlinear (quadratic, second-derivative)
destabilizing term.

The EE (\ref{eee10}) is relatively easy for numerical simulations of the 3D
waves in the inclined film. We have obtained good agreement with transient
states and transitions observed in the physical experiments
of Ref. \cn{ls95}.
Under certain parametric
conditions for which the dissipative terms of the EE are small, we
observed self-organization (from the initial white-noise small-amplitude
conditions) of unusual highly-ordered 
patterns of soliton-like structures on
the film surface (the pattern consists of two travelling-wave subpatterns
which move with different velocities). The studies of the evolution equation (%
\ref{eee10}) will be published elsewhere (Ref. \cn{if97}; see also 
Refs. \cn{fi96,if95}). 

A similar analysis leads to analogous
single-EE theory of a film flowing down a vertical ``fiber''. We believe
that similar theories can be useful for a variety of other systems.

\section{Acknowledgment}
This work is partially funded by the Department of Energy
Office of Basic Energy Sciences.
 
\appendix
\section{Validity conditions}
\label{aa1} 
From (\ref{ew1}), the estimate of $w_1$, in terms of the basic
parameters and length- and amplitude-scales, is 
\[
{w}_1\sim \mbox{max }\left( 
\frac{1}{Z^2},
~\frac{1}{Y^2},
~\frac RZ, 
~\frac{\cot\theta}{Z},
~\right .
\]
\begin{equation}
\left .
~\frac RT,
~\frac{W}{Z^3},
~\frac{W}{ZY^2} 
\right) A.  
\label{eomw1}
\end{equation}
Estimating $u_1$ from (\ref{eu1}), we have 
\[
{u}_1\sim \mbox{max }\left(
\frac{1}{Z^3}, 
~\frac{1}{ZY^2},
~\frac{R}{Z^2}, 
~\frac{\cot\theta}{Z^2},
~\frac{\cot\theta}{Y^2},
\right. 
\]
\begin{equation}
\left. 
\frac{R}{TZ},
~\frac{W}{Z^4},
~\frac{W}{Z^2Y^2},
~\frac{W}{Y^4}
\right) A.  \label{eomu1}
\end{equation}
In obtaining a solvable equation for the boundary condition on $w_{1x}$ at $%
x=1$ [Eq. (\ref{ebcw12})], we have dropped the term $u_{1z}(x=1)$. This
implies 
\begin{equation}
u_{1z}\ll w_{1x} ~~~ (x=1).  \label{euw1}
\end{equation}
Using the $O_M$ estimates for $u_1$, Eq. (\ref{eomu1}),
and $w_{1x}(x=1)$, Eq. (\ref
{ebcw12}), the above requirement reduces to 
\[
u_{1z}(x=1)\sim \mbox{max}  
\left (
~\frac{1}{Z^3}, 
~\frac{1}{ZY^2}, 
~\frac{R}{Z^2}, 
~\frac{\cot\theta}{Z^2},
~\frac{\cot\theta}{Y^2},
~\frac{R}{TZ},
\right . 
\]
\begin{equation}
\left .
~\frac{W}{Z^4}, 
~\frac{W}{Y^2Z^2},
~\frac{W}{Y^4} 
\right )
\left (\frac AZ\right )
\ll w_{1x}(x=1)\sim \frac{A}{Z^2}. \label{evcuw}
\end{equation}
This again yields the conditions (\ref{evc1y}), (\ref{evc1z}), (\ref{evc2}),
(\ref{evc3}), and, in addition, the following conditions:
\begin{equation}
\frac{\cot\theta}{Z}\ll 1,  \label{evc9}
\end{equation}
\begin{equation}
\left (\frac ZY\right )
\left (\frac{\cot\theta}{Y}\right )\ll 1  \label{evc11}
\end{equation}
\begin{equation}
\frac{W}{Z^3}\ll 1,  \label{evc8}
\end{equation}
\begin{equation}
\frac{W}{ZY^2}\ll 1,  \label{evc10}
\end{equation}
and
\begin{equation}
\left (\frac ZY\right )
\left (\frac{W}{Y^3}\right )\ll 1.  \label{evc12}
\end{equation}
These conditions, along with (\ref{evc0}), (\ref{evc1y}), (\ref{evc1z}), (\ref
{evc2}), and (\ref{evc3}), form the complete set of validity conditions for
the present theory. They are sufficient to justify all the simplifications
of the equations. Thus, the complete set of VCs is
\[
\mbox{max}\left[ 
A, 
~\frac{1}{Z^2},
~\frac{1}{Y^2},
~\frac RZ,
~\frac{\cot\theta}{Z},
~\frac{Z\cot\theta}{Y^2},
~\frac RT, 
\right .
\]
\begin{equation}
\left .
\frac{W}{Z^3}, 
~\frac{W}{Y^2Z}, 
~\left (\frac ZY\right )
\frac{WZ}{Y^4}
\right] \ll 1.  \label{evct}
\end{equation}
Using these conditions, it is easy to see that $w_1\ll w_0$ and $u_1\ll u_0$.

From (\ref{ep1}), we can estimate the $O_M$ ($Rp_1$):
\[
Rp_1\sim \mbox{max }\left( 
~\frac AZ,
~\frac{1}{Z^3}, 
~\frac{1}{ZY^2},
~\frac{R}{Z^2}, 
~\frac{\cot\theta}{Z^2},
~\frac{\cot\theta}{Y^2},
\right. 
\]
\begin{equation}
\left. 
\frac{R}{TZ}, 
~\frac{W}{Z^4}, 
~\frac{W}{Z^2Y^2},
~\frac{W}{Y^4}
\right) A.
\label{eomp1}
\end{equation}
Using the conditions (\ref{evct}), it is easy to show that $Rp_1/Z\ll Rp_0/Z$
or $p_1\ll p_0$. From (\ref{ev1}), we estimate the $O_M$ ($v_1$) as 
\[
v_1\sim \mbox{max }\left[ 
\frac{A}{YZ},
~\frac{1}{YZ^3}, 
~\frac{1}{ZY^3},
~\frac{R}{YZ^2},
~\frac{\cot\theta}{YZ^2}, 
\right. 
\]
\[
~\frac{\cot\theta}{Y^3},
~\frac{R}{TYZ}, 
~\frac{W}{YZ^4}, 
~\frac{W}{Y^3Z^2}, 
~\frac{W}{Y^5},
~\frac{A\cot \theta}{Y}, 
\]
\[
~\frac{AW}{YZ^2},
~\frac{AW}{Y^3},
~\frac{R\cot\theta}{ZY},
~\frac{RW}{YZ^3},
\]
\begin{equation}
\left. 
~\frac{RW}{ZY^3}, 
~\frac{R\cot\theta}{TY},
~\frac{RW}{TY^2Z},
~\frac{RW}{TY^3}
\right] A.  \label{eomv1}
\end{equation}
Using the estimate of $v_0$ [Eq. (\ref{eomv0})] and the conditions (\ref
{evct}), it is easy to see that $v_1\ll v_0$. The VCs (\ref{evct}) guarantee
that all the terms involving $w_0$, $u_0$, $v_0$, $p_0$, $w_1$, $u_1$, $v_1$%
, and $p_1$ that were dropped in obtaining solvable ODEs for the same
quantities are small in comparison with the terms that were retained.

Estimating the $O_M$ of various terms in Eq. (\ref{eee10}), and noting that $%
\delta \ll R$ or $\delta \sim R$, we find that 
\[
\frac AT\sim 
\mbox{max }\left [
A,
~\frac{1}{Z^2},
~\frac{1}{Y^2}, 
~\frac{\delta}{Z}, 
~\frac{\cot\theta}{Z}, 
~\frac{Z\cot\theta}{Y^2},
\right .
\]
\begin{equation}
\left .
~\frac{W}{Z^3},
~\frac{W}{ZY^2},
~\frac{WZ}{Y^4} 
\right ]
\left (\frac AZ\right )
\label{evrt1}
\end{equation}
and, consequently, taking (\ref{evct}) into account, 
\[
\frac RT \sim 
\mbox{max }\left [
~A,
~\frac{1}{Z^2},
~\frac{1}{Y^2}, 
~\frac{\delta}{Z},
~\frac{\cot\theta}{Z}, 
~\frac{\cot\theta Z}{Y^2}, 
\right .
\]
\begin{equation}
\left .
~\frac{W}{Z^3},
~\frac{WZ}{Y^4}, 
~\frac{W}{ZY^2} 
\right ]
\left (\frac RZ\right )
\ll 1.  \label{evrt}
\end{equation}
Hence, the parameter $R/T$ is small as a consequence of the smallness of
other parameters in (\ref{evct}), and thus can be omitted from there. Also, (%
\ref{ercot}) yields $\cot \theta /Z<R/Z\ll 1$, so that the parameter $\cot
\theta /Z$ can be omitted as well. The somewhat simplified validity
conditions are 
\[
\mbox{max}\left[ 
A, 
~\frac{1}{Z^2},
~\frac{1}{Y^2}, 
~\frac RZ, 
~\frac{\cot\theta Z}{Y^2},
\right .
\]
\begin{equation}
\left .
~\frac{W}{Z^3},
~\frac{W}{Y^2Z}, 
~\frac{WZ}{Y^4}
\right] \ll 1  \label{evctl}
\end{equation}
(which, in turn, significantly simplify [see Eq. (\ref{etvcn})]
if $Y\ge Z$). 

\section{Second and third iterative corrections}
\label{ab1} 
To find the second and third corrections for the velocity
components and pressure, we need first 
to determine $p_1$ and $v_1$, the
approximations to the exact solutions $\tilde{p}_1$ and $\tilde{v}_1$.
Considering the $x-$NS equation, 
\[
\tilde{p}_{1x}=-p_{0x}+\frac 1R (u_0+u_1)_{xx}
+\frac 1R \nabla^2 (u_0
+u_1)
\]
\[
-(u_0+u_1)_t 
-(u_0+u_1)(u_{0}+u_{1})_x -v_0(u_{0}+u_{1})_y 
\]
\[
-(w_N+w_0+w_1)(u_{0}+u_{1})_z 
+2(u_{0}+u_{1})_z 
\]
\begin{equation}
+\mbox{[terms containing $\tilde{v}_1$, $\tilde{w}_2$, or
$\tilde{u}_2$]}  \label{edep10}
\end{equation}
where 
\begin{equation}
\tilde{u}_2=\tilde{u}_1-u_1.  \label{etu1}
\end{equation}
Using our standard procedure, the simplified equation is 
\[
p_{1x}=\frac 1R u_{1xx}+\frac 1R \nabla^2u_0-w_Nu_{0z}+2u_{0z}-u_{0t} 
\]
\[
=\frac 2R (x-1)(W\nabla^4\eta-\cot\theta\nabla^2\eta)+\eta_{tz} 
\]
\begin{equation}
-\left (x^4-\frac 43 x^3+\frac 43\right )\eta_{zz}+(x^2+2x-5)
\frac{\nabla^2\eta_z}{R}
\label{edep11}
\end{equation}
where Eqs. (\ref{epn}) and (\ref{ep0}) for $p_N$ and $p_0$ have been
utilized. The normal-stress balance condition at $x=h$ yields 
\widetext
\[
\tilde{p}_1=-(p_N+p_0)+\frac 2R\left \{(u_0+u_1)_x+ v_{0y}\eta_y^2+
(w_{0}+w_{1})_z\eta_z^2 
-[(u_{0}+u_{1})_y +v_{0x}]\eta_y +[v_{0z}+(w_{0}+w_{1})_y]\eta_y\eta_z 
\right .
\]
\[
\left.
-[(u_{0}+u_{1})_z +w_{1x}]\eta_z\right \}
\{1+\eta_y^2+\eta_z^2\}^{-1} 
-\sigma \left[ \eta _{yy}\left( 1+\eta _z^2\right) +\eta _{zz}\left( 1+\eta
_y^2\right)  
-2\eta _y\eta _z\eta _{yz}\right] (1+\eta _y^2+\eta
_z^2)^{-3/2}  
\]
\begin{equation}
+\mbox{[terms containing $\tilde{v}_1$, $\tilde{w}_2$,
or $\tilde{u}_2$]} ~~~ (x=h)  
\label{ebcp10}
\end{equation}
\narrowtext
\noindent which simplifies to the BC 
\[
p_1=-p_{0x}\eta+\frac 2R u_{0xx}\eta+\frac 2R u_{1x} 
\]
\[
=-\frac 2R (W\nabla^4\eta-\cot\theta\nabla^2\eta) -\frac 53\eta_{zz} 
\]
\begin{equation}
-\frac{20}{3R}\nabla^2\eta_z-\frac 2R \eta\eta_z
+\frac 43 \eta_{tz} ~~~ (x=1).
\label{ebcp11}
\end{equation}
The solution is 
\[
Rp_1=(x^2-2x-1)\left [W\nabla^4\eta-\cot\theta\nabla^2\eta \right ] 
\]
\[
+\left (\frac{x^4}{3}-\frac{x^5}{5}
-\frac 43 x-\frac{7}{15}\right )R\eta_{zz} 
\]
\[
+\left (\frac{x^3}{3}+x^2-5x-3\right ) \nabla^2\eta_z 
\]
\begin{equation}
+\left (x+\frac 13\right )R\eta_{tz} -2\eta\eta_z  \label{ep1}
\end{equation}

The $y-$NS equation is 
\[
\tilde{v}_{1xx}=-v_{0xx}+R(p_N+p_0+p_1)_y
-\nabla^2 v_{0} +Rv_{0t} 
\]
\[
+R(u_0+u_1)v_{0x} 
+Rv_0v_{0y}+R(w_N+w_0+w_1-2)v_{0z} 
\]
\begin{equation}
+%
\mbox{[terms containing
$\tilde{w}_2$, $\tilde{u}_2$, $\tilde{p}_2$,
or $\tilde{v}_1$]},  \label{edev10}
\end{equation}
where $\tilde{p}_1=p_1+\tilde{p}_2$, and $v_0$ is given by
(\ref{ev0}). The simplified equation (for the approximation 
$v_1$) is 
\[
v_{1xx}=Rp_{1y}-\nabla^2 v_0 +R(w_N-2)v_{0z}+Rv_{0t} 
\]
or, in terms of $\eta$,
\widetext
\[
v_{1xx}=-(2x^2-4x-1)\cot\theta\nabla^2\eta_y 
+\left (\frac{2}{15}x^5+
\frac 23 x^4-\frac{16}{3}x^3+10x^2-\frac{28}{3}x
-\frac{7}{15}\right )R\eta_{zzy} 
\]
\[
-(x^4-4x^3+6x^2-4x)R\cot\theta\eta_{yz} 
+(x^2-2x)R\cot\theta\eta_{yt} -\left (\frac{x^3}{3}
+x^2-5x-\frac 13\right )R\eta_{zyt} 
\]
\begin{equation}
+P_{1}(\eta\eta_z)_y+P_{2}W\nabla^4\eta_y +P_{3}RW\nabla^2\eta_{yz} 
+P_{4}RW\nabla^2\eta_{yz}+P_{5}RW\nabla^2\eta_{yt},
\label{edev11}
\end{equation}
\narrowtext
\noindent where $v_1$ is the approximation to 
the exact solution $\tilde{v}_1$. The
coefficients of terms that are likely to contribute to spatial derivatives
of order five or more,
as well as the coefficients of nonlinear terms which
are small or dispersive in nature, are represented by $P_{1},~ \cdots,~
P_{5}$, which are polynomials in $x$; we will not need their explicit form,
just the fact that they are $\sim $ $1$. Henceforth, coefficients 
of type 
$P_{m}$ will always refer to such polynomials in $x$.
The boundary condition at $x=h$ is: 
\[
\tilde{v}_{1x}=-v_{0x}-(u_{0}+u_{1})_y+\left \{2[v_{0y}
-(u_{0}+u_{1})_x]\eta_y\right . 
\]
\[
+[(u_{0}+u_{1})_z+w_{1x}]\eta_y\eta_z 
\]
\[
\left . +[v_{0z}+(w_0+w_1)_y]\eta_z \right \}[1-\eta_y^2]^{-1} 
\]
\begin{equation}
+\mbox{[terms containing $\tilde{v}_1$, $\tilde{w}_2$, or
$\tilde{u}_2$]} ~~~ (x=h),  \label{ebcv10}
\end{equation}
where we have taken into account (\ref{ewn0}). The simplified BCs are 
\[
v_{1x}=-v_{0xx}\eta-u_{0yx}\eta-u_{1y}-2u_{0x}\eta_y+w_{0y}\eta_z 
\]
\[
=-\frac 23 W\nabla^4\eta_y-\frac{5}{12}R\eta_{tzy} +\frac{8}{15}R\eta_{zzy} 
\]
\[
+C_{1}(\eta\eta_z)_y
+C_{2}(\cot\theta)\eta\eta_y+C_{3}W\eta\nabla^2\eta_y 
\]
\begin{equation}
+C_{4}W\nabla^{4}\eta_y+C_{5}\nabla^{2}\eta_{yz} ~~~~~ (x=1)  
\label{ebcv11}
\end{equation}
and 
\begin{equation}
v_1=0 ~~~ (x=0)  \label{ensv1}
\end{equation}
where $C_{1},~ \cdots, ~ C_{5}$ are constants. (Henceforth, coefficients of
type $C_{m}$ are constants.) The solution is 
\widetext
\[
v_1=-\left (\frac{x^4}{6}-\frac{2}{3}x^3
-\frac{x^2}{2}+3x\right ) \cot\theta\nabla^2\eta_y 
+\left (\frac{x^7}{315}+\frac{x^6}{45}
-\frac{4x^5}{15}+\frac{5x^4}{6}-\frac{14}{9}x^3
-\frac{7x^2}{30}+\frac{158x}{45}\right )R\eta_{zzy} 
\]
\[
+\left (-\frac{x^6}{30}+\frac{x^5}{5}
-\frac{x^4}{2}+\frac 23 x^3-\frac 45 x\right ) R\cot\theta\eta_{yz} 
+\left (\frac{x^4}{12}
-\frac{x^3}{3}+\frac{2x}{3}\right )R\cot\theta\eta_{yt} 
\]
\[
-\left (\frac{x^5}{60}+\frac{x^4}{12}-\frac 56 x^3
-\frac{x^2}{6}+2x\right )R\eta_{tzy} 
+P_{6}(\eta\eta_z)_y +P_{7}(\cot\theta)\eta\eta_y
+P_{8}W\eta\nabla^2\eta_y 
\]
\begin{equation}
+P_{9}W\nabla^4\eta_y 
+P_{10}\nabla^2\eta_{yz} +P_{11}RW\nabla^2\eta_{yz}
+P_{12}RW\nabla^2\eta_{yt}.  \label{ev1}
\end{equation}
\narrowtext
By using the same procedure as above, we proceed to calculate $\tilde w%
_2$, $\tilde u_2$, $\tilde p_2$, and $\tilde v_2$. The equation for 
$\tilde{w}_2$ is
\[
\tilde w_{2xx}=-(w_0+w_1)_{xx}+R(p_0+p_1)_z
-\nabla ^2(w_0+w_1)+R(w_0+w_1)_t 
\]
\[
+R(u_0+u_1)(w_N+w_0+w_1)_x+R(v_0+v_1)(w_0+w_1)_y 
\]
\[
+R(w_N+w_0+w_1-2)(w_0+w_1)_z 
\]
\begin{equation}
+\mbox{[terms containing $\tilde{w}_2$, 
$\tilde{u}_2$, $\tilde{v}_2$, 
or $\tilde{p}_2$]}. \label{edew20}
\end{equation}
This yields 
\[
w_{2xx}=Rp_{1z}-\nabla ^2w_1+Rw_{1t}+Ru_1w_{Nx}+R(w_N-2)w_{1z} 
\]
\[
+Ru_0w_{0x}+Rw_0w_{0z}.
\]
Hence,
\widetext
\[
w_{2xx}=(1+4x-2x^2)\cot \theta \nabla ^2\eta _z 
-\left (\frac{x^5}{3}
+\frac{x^4}{2}-\frac{19x^3}{3}+5x^2
+\frac{4x}{3}\right )R\nabla ^2\eta _z 
\]
\[
+\left (\frac{7x^5}{15}-\frac{17x^3}{3}+12x^2
-\frac{34x}{3}-\frac{7}{15}\right )R\eta _{zzz} 
-(x^4-4x^3+6x^2-4x)\cot \theta \eta _{zz} 
\]
\[
+\left (-\frac{x^6}{10}+\frac{3x^5}{5}
-\frac{4x^4}{3}+\frac{4x^3}{3}+\frac{4x^2}{3}-\frac{8x}{3}
\right )R^2\eta _{zz} 
+\left (\frac{x^5}{3}-\frac{2x^4}{3}
-\frac{x^3}{3}+2x^2-2x\right )R\eta _{yyz} 
\]
\[
+\left (\frac{2x^4}{3}-\frac{8x^3}{3}
+2x^2\right )R\cot \theta \nabla ^2\eta 
-\left (\frac{x^5}{6}-\frac{2x^4}{3}+\frac{4x^3}{3}
+x^2-\frac{10x}{3}\right )R^2\eta _{tz} 
+(x^2-2x)R\cot \theta \eta _{tz}+2x^2R\eta \eta _z 
\]
\[
+P_{13}(\eta \eta _z)_z+P_{14}W\nabla ^4\eta _z+P_{15}\nabla ^2\eta
_{zz}+P_{16}\eta _{zzzz} 
+P_{17}\eta _{yyzz}+P_{18}RW\nabla ^2\eta _{zz}+P_{19}RW\nabla ^4\eta
+P_{20}R\nabla ^2\eta _t 
\]
\begin{equation}
+P_{21}RW\nabla ^2\eta _{tz}+P_{22}R^2\eta _{tt}+P_{23}R\eta
_{zzt}+P_{24}R\eta _{yyt}.  
\label{edew21}
\end{equation}
\narrowtext
\noindent The boundary condition at $x=h$ is 
\[
\tilde w_{2x}=-(w_0+w_1)_x-(u_0+u_1)_z+\{2[(w_0+w_1)_z-(u_0+u_1)_x]\eta _z 
\]
\[
+[(u_0+u_1)_y+(v_0+v_1)_x]\eta _y\eta _z 
\]
\[
+[(v_0+v_1)_z+(w_0+w_1)_y]\eta _y\}\{1-\eta _z^2\}^{-1} 
\]
\begin{equation}
+%
\mbox{[terms containing $\tilde{w}_2$, $\tilde{u}_2$,
$\tilde{v}_2$, or $\tilde{p}_2$]}~~~(x=h).  \label{ebcw20}
\end{equation}
Dropping all the unknown and smaller terms, we have, at $x=1$, 
\[
w_{2x}=-w_{1xx}\eta -u_{0zx}\eta -u_{1z}+2(w_{0z}-u_{0x})\eta _z+w_{0y}\eta
_y 
\]
\[
=\frac{8}{15}R\eta _{zzz}-\frac 23 \cot \theta \nabla ^2\eta _z
+2R\eta \eta _z 
-2\cot \theta \eta \eta _z+C_{6}W\eta \nabla ^2\eta _z
\]
\[
+C_{7}(\eta \eta
_z)_z+C_{8}(\eta \eta _y)_y 
+C_{9}W\nabla ^4\eta _z+C_{10}\nabla ^2\eta _{zz}
\]
\begin{equation}
+C_{11}R\eta _{tzz}+C_{12}R\eta _t\eta ~~~(x=1).  \label{ebcw21}
\end{equation}
Finally, the no-slip condition is 
\begin{equation}
w_2=0~~~(x=0).  \label{ensw2}
\end{equation}
Integrating (\ref{edew21}) with the boundary conditions (\ref{ebcw21}) and (%
\ref{ensw2}), we obtain 
\widetext
\[
w_2=\left (-\frac{x^4}{6}+\frac{2x^3}{3}
+\frac{x^2}{2}-3x\right )\cot \theta \nabla ^2\eta _z 
+\left (-\frac{x^7}{126}-\frac{x^6}{60}
+\frac{19x^5}{60}-\frac{5x^4}{12} 
-\frac{2x^3}{9}+\frac{163x}{180}\right )R\nabla ^2\eta _z 
\]
\[
+\left (\frac{x^7}{90}-\frac{17x^5}{60}
+x^4-\frac{17x^3}{9}-\frac{7x^2}{30}+\frac{721x}{180}\right 
)R\eta _{zzz} 
+\left (-\frac{x^8}{560}+\frac{x^7}{70}-\frac{2x^6}{45}
+\frac{x^5}{15}+\frac{x^4}{9} 
-\frac{4x^3}{9}+\frac{232x}{315}\right )R^2\eta _{zz} 
\]
\[
+\left (-\frac{x^6}{30}+\frac{x^5}{5}
-\frac{x^4}{2}+\frac{2x^3}{3}-\frac{4x}{5}\right
)R\cot \theta \eta _{zz} 
+\left (\frac{x^7}{126}-\frac{x^6}{45}
-\frac{x^5}{60}+\frac{x^4}{6}-\frac{x^3}{3}+\frac{89x}{180}
\right )R\eta _{yyz} 
\]
\[
+\left (\frac{x^6}{45}-\frac{2x^5}{15}
+\frac{x^4}{6}-\frac{2x}{15}\right )R\cot \theta \nabla ^2\eta 
+\left (-\frac{x^7}{252}+\frac{x^6}{45}
-\frac{x^5}{15}-\frac{x^4}{12}+\frac{5x^3}{9} 
-\frac{199x}{180}\right )R^2\eta _{tz}
\]
\[
+\left (\frac{x^4}{12}-\frac{x^3}{3}+\frac{2x}{3}\right 
)R\cot \theta \eta _{tz} 
+\left (\frac{x^4}{6}+\frac{4x}{3}\right
)R\eta \eta _z-2x\cot \theta \eta \eta _z 
+P_{25}(\eta \eta _z)_z+P_{26}W\eta \nabla ^2\eta _z
\]
\[
+P_{27}(\eta \eta _y)_y 
+P_{28}W\nabla ^4\eta _z+P_{29}\nabla ^2\eta _{zz}+P_{30}\eta
_{zzzz}+P_{31}\eta _{yyzz} 
+P_{32}RW\nabla ^2\eta _{zz}+P_{33}RW\nabla ^4\eta 
+P_{34}RW\nabla ^2\eta
_{tz}
\]
\begin{equation}
+P_{35}R^2\eta _{tt} 
+P_{36}R\eta _{zzt}+P_{37}R\eta _{yyt}+P_{38}R\nabla ^2\eta _t+P_{39}R\eta
\eta _t.  \label{ew2}
\end{equation}
\narrowtext
\noindent The equation for $\tilde 
u_2$, obtained from the incompressibility condition, is 
\begin{equation}
\tilde u_{2x}=-(v_1+\tilde v_2)_y
-(w_2+\tilde w_3)_z  \label{edeu20}
\end{equation}
where $\tilde w_2=w_2+\tilde w_3$. Dropping the unknown terms,
we obtain 
\[
{u}_{2x}=-v_{1y}-w_{2z} 
\]
or
\widetext
\[
u_{2x}=\left (\frac{x^6}{90}
-\frac{x^5}{15}+\frac{x^4}{3}-\frac{2x^3}{3}
+\frac{14x}{15}\right )R\cot \theta \nabla ^2\eta _z 
+\left (\frac{x^8}{560}-\frac{x^7}{70}
+\frac{2x^6}{45}-\frac{x^5}{15}-\frac{x^4}{9}+\frac{4x^3}{9} 
-\frac{232x}{315}\right )R^2\eta _{zzz}
\]
\[
+\left (\frac{x^4}{6}-\frac{2x^3}{3}-\frac{x^2}{2}
+3x\right )\cot \theta \nabla ^4\eta 
+\left (-\frac{x^7}{315}+\frac{x^6}{60}-\frac{x^5}{30}
-\frac{7x^4}{12}+\frac{19x^3}{9}+
\frac{7x^2}{30}-\frac{221x}{45}\right )R\nabla ^2\eta _{zz} 
\]
\[
-\left (\frac{x^4}{12}-\frac{x^3}{3}
+\frac{2x}{3}\right )R\cot \theta \nabla ^2\eta _t 
+\left 
(\frac{x^7}{252}-\frac{x^6}{45}
+\frac{x^5}{15}+\frac{x^4}{12}-\frac{5x^3}{9}+\frac{199x}{180}
\right )R^2\eta _{tzz} 
-\left (\frac{x^4}{6}+\frac{4x}{3}\right
)R(\eta \eta _z)_z
\]
\[
+2x\cot \theta (\eta \eta _z)_z 
+P_{39}(\eta \eta _z)_{zz}+P_{40}W(\eta \nabla ^2\eta _z)_z+P_{41}(\eta \eta
_y)_{yz} 
+P_{42}(\eta \eta _z)_{yy}+P_{43}(\eta _z\eta _y)_y+P_{44}W\nabla ^4\eta
_{zz} 
\]
\[
+P_{45}\nabla ^2\eta _{zzz}+P_{46}\eta _{zzzzz}+P_{47}\eta _{yyzzz} 
+P_{48}RW\nabla ^2\eta _{zzz}+P_{49}RW\nabla ^4\eta _z+P_{50}W\nabla ^4\eta
_{yy}+P_{51}\nabla ^2\eta _{zyy} 
\]
\[
+P_{52}RW\nabla ^2\eta _{zyy}
+P_{53}RW\nabla ^2\eta _{tyy}
+P_{54}R\eta_{tzyy}
+P_{55}RW\nabla^2\eta _{tzz}
+P_{56}R^2\eta _{ttz}+P_{57}R\eta _{tzzz} 
\]
\begin{equation}
+P_{58}R\eta _{tyyz}+P_{59}R\nabla ^2\eta _{tz}+P_{60}R(\eta \eta _t)_z
\label{edeu21}
\end{equation}
\narrowtext
\noindent where $u_2$ is the approximation to 
$\tilde u_2$. The no-slip boundary
condition is 
\begin{equation}
u_2=0 ~~~~ (x=0).  \label{ensu2}
\end{equation}
Integrating Eq. (\ref{edeu21}) with the no-slip BC, we
obtain 
\widetext
\[
u_2= 
\left (\frac{x^9}{5040}-\frac{x^8}{560}
+\frac{2x^7}{315}-\frac{x^6}{90}-\frac{x^5}{45} 
+\frac{x^4}{9}-\frac{116x^2}{315}\right )R^2\eta _{zzz}
\]
\[
+\left (\frac{x^7}{630}-\frac{x^6}{90}+\frac{x^5}{15}
-\frac{x^4}{6}+\frac{7x^2}{15}\right )R\cot \theta \nabla ^2\eta _z 
+\left (\frac{x^5}{30}-\frac{x^4}{6}
-\frac{x^3}{6}+\frac{3x^2}{2}\right )\cot \theta \nabla ^4\eta 
\]
\[
+\left (-\frac{x^8}{2520}+\frac{x^7}{420}
-\frac{x^6}{180}-\frac{7x^5}{60}+\frac{19x^4}{36} 
+\frac{7x^3}{90}-\frac{221x^2}{90}\right )R\nabla ^2\eta _{zz} 
\]
\[
+\left (\frac{x^8}{2016}-\frac{x^7}{315}
+\frac{x^6}{90}+\frac{x^5}{60}-\frac{5x^4}{36}
+\frac{199x^2}{360}\right )R^2\eta _{tzz}
-\left (\frac{x^5}{60}-\frac{x^4}{12}
+\frac{x^2}{3}\right )R\cot \theta \eta _{tzz} 
-\left (\frac{x^5}{30}
+\frac{2x^2}{3}\right )R(\eta \eta _z)_z
\]
\[
+x^2\cot \theta (\eta \eta _z)_z 
+P_{61}(\eta \eta _z)_{zz}+P_{62}(\eta \eta _y)_{yz}+P_{63}(\eta \eta
_z)_{yy}+P_{64}(\eta _z\eta _y)_y 
+P_{65}W(\eta\nabla^2\eta_z)_z
+P_{66}W\nabla ^4\eta _{zz}
\]
\[
+P_{67}\nabla ^2\eta_{zzz}
+P_{68}\eta _{zzzzz} 
+P_{69}\eta _{yyzzz}+P_{70}RW\nabla ^2\eta _{zzz}+P_{71}RW\nabla ^4\eta _z 
+P_{72}W\nabla ^4\eta _{yy}
\]
\[
+P_{73}\nabla ^2\eta _{zyy}
+P_{74}RW\nabla ^2\eta
_{zyy}+P_{75}RW\nabla ^2\eta _{tyy} 
+P_{76}R\eta _{tzyy}+P_{77}RW\nabla ^2\eta _{tzz}+P_{78}R^2\eta_{ttz}
+P_{79}R\eta _{tzzz} 
\]
\begin{equation}
+P_{80}R\eta _{tyyz}+P_{81}R\nabla ^2\eta _{tz}+P_{82}R(\eta \eta _t)_z.
\label{eu2}
\end{equation}
\narrowtext
\noindent In order to obtain all the fourth-order linear dissipative terms
of the EE, we have to
proceed to the next approximation
[since (see Appendix \ref{ac1}) the dissipative terms of $u_3$ are
not guaranteed to be much smaller than those of $u_2$; however,
the dissipative terms of $u_4$ {\it are} much smaller than those of
$u_2$]. Expressing $\tilde{u}_2=u_2+\tilde{u}_3$,
the incompressibility condition leads to an
equation for the approximation $u_3$:  
\begin{equation}
u_{3x}=-v_{2y}-w_{3z},  \label{edeu30}
\end{equation}
(where $v_2$ and $w_3$ are the approximations to $\tilde{v}_2$ 
and $\tilde{w}_3$
respectively). The solvable ODEs for $v_2$ and $w_3$
will contain $p_2$, 
the approximation to $\tilde{p}_2$. 
Terms from $p_2$ will not make any
contribution to linear terms with derivatives of order four or less
and the nonlinear terms in the final EE. 
However, in Appendix \ref{ac1},
we will obtain an equation for the general iterate $p_n$, and
the form of the $x-$NS equation for $p_2$ will guide us in formulating the
general equation for $p_n$.

The $x-$NS equation for pressure $\tilde{p}_2$ can be written
in the form 
\widetext
\[
\tilde{p}_{2x}=-(p_0+p_1)_x+\frac 1R (u_0+u_1+u_2)_{xx}
+\frac 1R \nabla^2(u_0+u_1+u_2)-(u_0+u_1+u_2)_t 
-(u_0+u_1+u_2)(u_0+u_1+u_2)_x 
\]
\[
-(v_0+v_1)(u_0+u_1+u_2)_y 
-(w_N+w_0+w_1+w_2-2)(u_{0}+u_{1}+u_{2})_z 
\]
\begin{equation}
+\mbox{ 
[terms containing
$\tilde{v}_2$, $\tilde{w}_3$, or $\tilde{u}_3$ ]}  \label{edep20}
\end{equation}
\narrowtext
\noindent which yields 
\[
p_{2x}=-p_{1x}\eta+\frac{u_{2xx}}{R}
+\frac{\nabla^2u_1}{R}-u_{1t}-2u_{1z} 
\]
\begin{equation}
-w_Nu_{1z}-u_0u_{0x}-w_0u_{0z}.  \label{edep21}
\end{equation}
The boundary condition on pressure at $x=h$ is 
\widetext
\[
\tilde{p}_2=-(p_N+p_0+p_1)+
\frac 2R\{(u_0+u_1+u_2)_x+(v_{0}+v_{1})_y\eta_y^2 
+(w_0+w_1+w_2)_z\eta_z^2 
-[(u_0+u_1+u_2)_y+(v_0+v_1)_x]\eta_y 
\]
\[
+[(v_0+v_1)_z+(w_0 +w_1+w_2)_y]\eta_y\eta_z 
-[(u_0 +u_1+u_2)_z+(w_1+w_2)_x]\eta_z\}
(1+\eta_y^2+\eta_z^2)^{-1} 
\]
\[
-\sigma \left[ \eta _{yy}\left( 1+\eta _z^2\right) +\eta _{zz}\left( 1+\eta
_y^2\right) \right. 
\left. -2\eta _y\eta _z\eta _{yz}\right] (1+\eta _y^2+\eta
_z^2)^{-3/2}   
\]
\begin{equation}
+\mbox{[terms containing
$\tilde{v}_2$, $\tilde{w}_3$, or $\tilde{u}_3$]} ~~~ (x=h).
\label{ebcp20}
\end{equation}
\narrowtext
\noindent This yields, at $x=1$, 
\[
p_2=-p_{1x}\eta+[\frac 2R (u_{1xx}\eta+u_{2x}-(u_{0y}+v_{0x})\eta_y 
\]
\[
-(u_{0z}+w_{1x})\eta_z] 
-\sigma[\eta_{yy}\eta_z^2+\eta_{zz}\eta_y^2-2\eta_y\eta_z\eta_{yz}]
\]
\begin{equation}
-(3/2)\sigma\nabla^2\eta(\eta_y^2+\eta_z^2) ~~~ (x=1).  \label{ebcp21}
\end{equation}
As mentioned before, $p_2$ will not contribute to linear terms with
derivatives of order 4 or less and the nonlinear terms. Hence, we will
not solve for $p_2$.

Representing $\tilde{p}_2$ as $\tilde{p}_2=p_2+\tilde{p}_3$, the $y-$NS
equation for $\tilde{v}_2$ is 
\widetext
\[
\tilde{v}_{2xx}=-(v_0+v_1)_{yy}+R(p_0+p_1+p_2)_y
-\nabla^2(v_0+v_1)+R(v_0+v_1)_t 
+R(u_0+u_1+u_2)(v_{0}+v_{1})_x+R(v_0+v_1)(v_{0}+v_{1})_y 
\]
\begin{equation}
+R(w_N+w_0+w_1+w_2-2)(v_{0}+v_{1})_z 
+\mbox{[terms containing $\tilde{v}_2$, $\tilde{w}_3$, $\tilde{u}_3$,
or $\tilde{p}_3$]}  
\label{edev20}
\end{equation}
\narrowtext
\noindent The simplified equation for the approximation $v_2$ is 
\[
{v}_{2xx}=Rp_{2y}-\nabla^2v_1+Rv_{1t} 
\]
\begin{equation}
+Ru_0v_{0x}+R(w_N-2)v_{1z}+Rw_0v_{0z}.  \label{edev21}
\end{equation}
The boundary condition on $\tilde{v}_2$ at $x=h$ is 
\widetext
\[
\tilde{v}_{2x}=-(v_{0}+v_{1})_x-(u_0+u_1+u_2)_y 
+\{2[(v_0+v_1)_y-(u_0+u_1+u_2)_x]\eta_y 
+[(u_0+u_1+u_2)_z+(w_1+w_2)_x]\eta_y\eta_z 
\]
\begin{equation}
+[(v_0+v_1)_z+(w_0+w_1+w_2)_y]\eta_z\}(1-\eta_y^2)^{-1} 
+\mbox{[terms containing
$\tilde{v}_2$, $\tilde{w}_3$, or $\tilde{u}_3$]} ~~~
(x=h).  \label{ebcv20}
\end{equation}
\narrowtext
\noindent The simplified BC on $v_2$ at $x=1$ is 
\[
v_2=-\frac 12 v_{0xxx}\eta^2-v_{1xx}\eta-
(\frac 12 u_{0yxx}\eta^2+u_{1yx}\eta+u_{2y}) 
\]
\[
+[2(v_0-u_{0xx}\eta-u_{1x})\eta_y+ 
\]
\begin{equation}
+(v_{0z}+w_{0yx}\eta+w_{1y})\eta_z] ~~~~~ (x=1).  \label{ebcv21}
\end{equation}
The no-slip condition on $v_2$ is 
\begin{equation}
v_2(x=0)=0.  \label{ensv2}
\end{equation}
As mentioned before, we will not solve the equation for $v_2$ to find all the
terms in it. We are only interested in the terms that make
linear dissipative contributions of order four or less to the
EE. Only the term involving $R(w_N-2)v_{1z}$ makes such a
contribution. There is no contribution from
the tangential stress and the no-slip conditions. Thus,
\[
v_{2xx}^{(\le 3)}=R(w_N-2)v_{1z} 
\]
\[
=\left (\frac{x^8}{30}-\frac{4x^7}{15}
+\frac{29x^6}{30}-\frac{31x^5}{15}+\frac{7x^4}{3} \right .
\]
\begin{equation}
\left . -\frac{8x^3}{15}-\frac{8x^2}{5}
+\frac{8x}{5}\right )R^2\cot\theta\eta_{yzz}  \label{edev22}
\end{equation}
where the superscript on $v_{2xx}$ indicates that 
RHS of the above expression
contains only
linear terms with spatial derivatives of order $3$ or less with all
other terms omitted.
 
\noindent Solving for $v_2^{(\le 3)}$, we obtain 
\[
v_2^{(\le 3)}= 
\left (\frac{x^{10}}{2700}-\frac{x^9}{270}
+\frac{29x^8}{1680}-\frac{31x^7}{630}
+\frac{7x^6}{90}
\right .
\]
\begin{equation}
\left .
-\frac{2x^5}{75}-\frac{2x^4}{15}+\frac{4x^3}{15}
-\frac{344x}{945}\right )R^2\cot\theta\eta_{zzy}.  \label{ev2}
\end{equation}

Similarly, proceeding to $w_3$, we find that only two terms in the 
$z-$NS equation make contributions to fourth derivative terms
in $u_3$, i.e. 
\[
w_{3xx}^{(\le 3)} 
=Rw_{Nx}u_2+R(w_N-2)w_{2z},
\]
or
\widetext
\[
w_{3xx}^{(\le 3)} 
=\left (\frac{x^{10}}{720}
-\frac{x^9}{72}+\frac{19x^8}{315}
-\frac{47x^7}{315}+\frac{2x^6}{15} 
+\frac{4x^5}{15}-\frac{8x^4}{9}
+\frac{8x^3}{9}+\frac{232x^2}{315}-\frac{464x}{315}\right 
)R^3\eta_{zzz} 
\]
\[
+\left (-\frac{8x^8}{315}
+\frac{64x^7}{315}-\frac{19x^6}{30}+\frac{16x^5}{15}
-\frac{2x^4}{3} 
-\frac{4x^3}{5}-\frac{4x^3}{5}
+\frac{2x^2}{3}+\frac{4x}{15}
\right )R^2\cot\theta\nabla^2\eta_z  
\]
\begin{equation}
+\left (\frac{x^8}{30}-\frac{4x^7}{15}
+\frac{29x^6}{30}-\frac{31x^5}{15}+\frac{7x^4}{3}
-\frac{8x^3}{15}-\frac{8x^2}{5}
+\frac{8x}{5}\right )R^2\cot\theta\eta_{zzz}.
\label{edew30}
\end{equation}
\narrowtext
\noindent We note that only the terms with 
$R\cot\theta\nabla^2\eta_z$ and $%
R^2\eta_{zzz}$ in $u_2$ [see Eq. (\ref{eu2})] make a contribution. Similarly,
only the terms with $R\cot\theta\eta_{zz}$, $R^2\eta_{zz}$, and $%
R\cot\theta\nabla^2\eta$ in $w_2$ [see Eq. (\ref{ew2})] make the relevant
contribution. Again, there are no contributions from the tangential and
no-slip BCs.

The relevant part of $w_3$ is 
\widetext
\[
w_3^{(\le 3)}=
\left (\frac{x^{12}}{95040}-\frac{x^{11}}{7920}
+\frac{19x^{10}}{28350}-\frac{47x^9}{22680}
+\frac{x^8}{420}+\frac{2x^7}{315}
-\frac{4x^6}{135}
\right .
\]
\[
\left .
+\frac{2x^5}{45}
+\frac{58x^4}{945}-\frac{232x^3}{945}
+\frac{12358x}{31185}\right )R^3\eta _{zzz}
\]
\[
+\left (\frac{x^{10}}{2700}-\frac{x^9}{270}
+\frac{29x^8}{1680}-\frac{31x^7}{630}+\frac{7x^6}{90}
-\frac{2x^5}{75}-\frac{2x^4}{15}
+\frac{4x^3}{15}-\frac{344x}{945}\right )R^2\cot \theta \eta _{zzz}
\]
\begin{equation}
+\left (-\frac{4x^{10}}{14175}+\frac{8x^9}{2835}
-\frac{19x^8}{1680}+\frac{8x^7}{315}-\frac{x^6}{45}
-\frac{x^5}{25}+\frac{x^4}{18}
+\frac{2x^3}{45}-\frac{107x}{810}\right 
)R^2\cot \theta \nabla ^2\eta _z.  \label{ew3}
\end{equation}
\noindent \narrowtext
Using the expressions for $v_2$ and $w_3$ 
in the incompressibility condition 
(\ref{edeu30}), we obtain 
\widetext
\[
u_3^{(\le 4)}=
-\left (\frac{x^{13}}{1235520}
-\frac{x^{12}}{95040}+\frac{19x^{11}}{311850}-\frac{47x^{10}}{226800}
\right .
\]
\[
\left .
+\frac{x^9}{3780}
+\frac{x^8}{1260}-\frac{4x^7}{945}
+\frac{x^6}{135}+\frac{58x^5}{4725}
-\frac{58x^4}{945}+\frac{6179x^2}{31185}\right
)R^3\eta _{zzzz}
\]
\begin{equation}
-\left (\frac{x^{11}}{124740}-
\frac{x^{10}}{11340}+\frac{x^9}{1512}-\frac{x^8}{336}
+\frac{x^7}{126}-\frac{x^6}{90}
-\frac{7x^5}{450}+\frac{7x^4}{90}
-\frac{2813x^2}{11340}\right 
)R^2\cot \theta \nabla ^2\eta _{zz}.  \label{eu3}
\end{equation}
\narrowtext
\noindent Following our standard procedure 
[see Eqs. (\ref{ekc10}) and (\ref{ekc1})],
the kinematic condition at $x=1$ is written as 
\[
\eta _t+(w_0+w_1)\eta _z
\]
\begin{equation}
=u_{0x}\eta +u_1+u_{1x}\eta +u_2+u_3~~~(x=1)   \label{ekc2}
\end{equation}
where we have taken into account Eq. (\ref{ekc01}). 
Also, terms $w_1\eta_z$ and $u_{1x}\eta$ which could normally be
discarded in comparison with $u_1$, must be kept because $u_1(x=1)$
contains $\eta_{xx}$ with the factor $\delta$ which is allowed to be 
small.
Note that we have not shown terms of type $v_0\eta _y$, $v_1\eta _y$, $%
v_2\eta _y$ and $w_2\eta _z$ as they make smaller contributions. Also, $%
w_{Nx}\eta =0$ at $x=1$. Using the expressions for the velocities (\ref{ew0}%
), (\ref{ew1}), (\ref{eu0}), (\ref{eu1}), (\ref{eu2}), and (\ref{eu3}) in 
Eq. (\ref{ekc2}), we obtain
\widetext
\[
\left[ \eta -\frac{5}{12}R\eta _z+\frac{4}{15}R\cot \theta \nabla ^2\eta
-\frac{295}{672}R^2\eta _{zz}\right ]_t
+4\eta \eta _z+\frac{8}{15}R\eta _{zz}
-\frac 23\cot \theta \nabla ^2\eta 
\]
\[
+2\nabla
^2\eta _z+\frac 23 W\nabla ^4\eta 
+\left [\frac{23}{15} R-2\cot \theta \right ]
(\eta \eta _z)_z
-\frac{5}{14}R\cot \theta \nabla ^2\eta
_z
+\frac 27 R^2\eta _{zzz}-\frac 65
\cot \theta \nabla ^4\eta 
\]
\begin{equation}
+\frac{331}{168}R\nabla ^2\eta
_{zz}
+\frac{1241483}{8108100}R^3\eta _{zzzz}
-\frac{477523}{2494800}
R^2\cot \theta \nabla ^2\eta _{zz}=0. 
\label{eee23a}
\end{equation}
\narrowtext
\noindent In this equation, we do not 
shown the linear terms with 
derivatives of order greater than four. 
Also, we have not shown the nonlinear terms with the
derivatives of odd orders, of
the type $\eta _z\eta _{yy}$, $\eta _y\eta _{zy}$, $(\eta
\eta _{zz})_z$, etc. Using a procedure identical to the one 
of Refs. \cn{os69,os70} (see also Ref.
\cn{ad90}), it can be shown that the
nonlinear terms with odd-order
derivatives are dispersive in nature, (i.e. they do
not give rise to any change in the amplitude). 
In comparison with the linear
dispersive term $\nabla ^2\eta _z$, the nonlinear dispersive terms are
small. Similarly, nonlinear terms with even-order
derivatives can be shown to be
dissipative. Among the nonlinear dissipative terms, the terms involving $W$%
 are of the type $W\eta \nabla ^4\eta $, $W\eta _y\nabla ^2\eta _y$, etc.
These terms are smaller than the linear dissipative term $\sim W\nabla
^4\eta ,$ $\cot \theta \eta _{yy},$ and 
$\delta \eta _{zz}$ by the factor $A$.
However, the nonlinear dissipative term $R(\eta \eta _z)_z$ may be $\sim
\delta \eta _{zz}$ when $\delta \ll R$.
\section{General analysis of iteration procedure}
\label{ac1} 
We want to analyze the general \textit{n}th step of the iteration procedure.
Since, as was pointed out in the main text, this is of interest mainly for
the near-critical conditions, $R\approx R_c=(5/4)\cot \theta $, where the
flow is essentially 2D, $Y\gg Z$, for simplicity we will consider from the
outset a 2D flow ($\partial _y=v=0$) with $\cot \theta =0$.

From the preceding iteration steps, the corrections are known up to
(including) $u_{n-1}$, $w_{n-1}$, and $p_{n-2}.$ We substitute into the exact
NS problem the expansion

\[
w=w_N+w_0+w_1+\cdot \cdot \cdot w_{n-1}+\widetilde{w_n} 
\]
where $\widetilde{w_n}$ is an unknown correction and all the other members
of the sum are known from the preceding iteration steps, and similar
expansions for $u$ and $p$. The exact equations and boundary conditions for $%
\widetilde{p_{n-1}}$ are 
\widetext
\[
R\tilde{p}_{(n-1)x}=-\sum_{k=0}^{n-2}Rp_{kx}+
\sum_{k=0}^{n-1}\left [u_{kxx}+u_{kzz}-Ru_{kt} 
-R(w_N-2)u_{kz}
\right ]
-R\left (\sum_{k=1}^{n-1}u_k\right )
\left (\sum_{k=1}^{n-1}u_{kx}\right )
\]
\begin{equation}
-R\left (\sum_{k=1}^{n-1}w_k\right )
\left (\sum_{k=1}^{n-1}u_{kz}\right )
+\mbox{[terms containing $\tilde{w}_n$ or $\tilde{u}_n$]}
\label{cccp}
\end{equation}
\narrowtext
\noindent and 
\widetext
\[
R\tilde{p}_{n-1}=-Rp_N-\sum_{k=0}^{n-2}Rp_k
-2\left 
(1-\eta_z^2+\eta_z^4+\cdots +(-1)^p\eta_z^{2p}
\right )
\sum_{k=0}^{n-1}u_{kx}
\]
\[
+2\left (\eta_z-\eta_z^3+\eta_z^5-\cdots \right )
\sum_{k=0}^{n-1}\left (u_{kz}+w_{kx}\right )
+2\left (\eta_z-\eta_z^3+\eta_z^5-\cdots \right )
w_{Nx}
\]
\begin{equation}
-2\left (\eta_z^2-\eta_z^4+\eta_z^6-\cdots \right )
\sum_{k=0}^{n-1} w_{kz}
+\mbox{[terms containing $\tilde{w}_n$
or $\tilde{u}_n$]}
~~~ (x=h),
\label{cccpb}
\end{equation}
and the corresponding equations for the approximate $p_{n-1}$ are written as 
\[
Rp_{(n-1)x}=u_{(n-1)xx}+u_{(n-2)zz}-Ru_{(n-2)t} 
-R(w_N-2)u_{(n-2)z} 
\]
\begin{equation}
-R\Sigma_{k=0,..,n-3}\left
[u_ku_{(n-3-k)x}
+w_ku_{(n-3-k)z}\right ]  \label{cccpa}
\end{equation}
and
\widetext
\[
Rp_{n-1}=-Rp_N^{(n)}\frac{\eta^n}{n!}
-\sum_{k=0}^{n-2}Rp_k^{(n-k-1)}\frac{\eta^{n-k-1}}{(n-k-1)!}
\]
\[
-\sum_{k=0}^{[(n-k-1)/3]}
\left [2(-1)^p\eta_z^{2p}
\sum_{k=0}^{n-3p-1}u_{kx}^{(n-k-3p-1)}\frac{\eta^{n-k-3p-1}}{(n-k-3p-1)!}
\right ]
\]
\[
+\sum_{p=0}^{[(n-k-3)/3]}\left [
2(-1)^p\eta_z^{2p+1}
\sum_{k=0}^{n-3p-3}u_{kz}^{(n-k-3p-3)}\frac{\eta^{n-k-3p-3}}{(n-k-3p-3)!}
\right ]
\]
\[
+\sum_{p=0}^{[(n-k-2)/3]}\left [
2(-1)^p\eta_z^{2p+1}
\sum_{k=0}^{n-3p-2}w_{kx}^{(n-k-3p-2)}\frac{\eta^{n-k-3p-2}}{(n-k-3p-2)!}
\right ]
\]
\[
-\sum_{k=0}^{[(n-k-4)/3]}\left [
2(-1)^p\eta_z^{2p+2}
\sum_{k=0}^{n-3p-4}w_{kz}^{(n-k-3p-4)}\frac{\eta^{n-k-3p-4}}{(n-k-3p-4)!}
\right ]
\]
\begin{equation}
+2\left ( \eta_z-\eta_z^3+\eta_z^5-\cdots \right )w_{Nx}
~~~~ (x=1)
\label{cccpba}
\end{equation}
\narrowtext
\noindent where the notation $[\cdots]$ above the summation sign $\sum$
indicates that only the integer part of the expression inside the
brackets needs to be considered.

After the solution $p_{n-1}$ has been found, the exact equations 
\[
\tilde{w}_{nxx}=\sum_{k=0}^{n-1}\left [-w_{kxx}-w_{kzz}
+R(w_N-2)w_{kz}
+Rw_{kt}
\right .
\]
\[
\left .
+Ru_{k}w_{Nx} 
+Rp_{kz}\right ]
\]
\[
+R\left (\sum_{k=0}^{n-1}u_k\right )
\left (\sum_{k=0}^{n-1}w_{kx}\right )
+R\left (\sum_{k=0}^{n-1}w_k\right )
\left (\sum_{k=0}^{n-1}w_{kz}\right )
\]
\begin{equation}
+\mbox{[terms containing $\tilde{w}_n$, $\tilde{u}_n$,
or $\tilde{p}_n$]},
\label{cccw}
\end{equation}
\[
\tilde{w}_{nx}=-w_{Nx}-\sum_{k=0}^{n-1}\left (
w_{kx}+u_{kz}\right )
\]
\[
+2\sum_{k=0}^{n-1}\left (w_{kz}\eta_z-u_{kx}\eta_z\right )
\left (1-\eta_z^2\right )^{-1}
\]
\begin{equation}
+\mbox{[terms containing $\tilde{w}_n$, $\tilde{u}_n$,
$\tilde{v}_n$ or $\tilde{p}_n$]} ~~~~~ (x=h),
\label{cccwb}
\end{equation} 
and
\begin{equation}  
\tilde{w}_n=0 ~~~~~ (x=0)
\label{cccw0}
\end{equation}
lead to the system for $w_n$, 
\[
w_{nxx}=
-w_{(n-1)zz}
+R(w_N-2)w_{(n-1)z}
+Rw_{(n-1)t} 
\]
\[
+Ru_{(n-1)}w_{Nx}
+Rp_{(n-1)z}
+R\Sigma_{k=0,..,n-2}\left
[u_kw_{(n-2-k)x}\right .
\]
\begin{equation}
\left .
+w_kw_{(n-2-k)z}\right ],  \label{cccwa}
\end{equation}
\widetext
\[
w_{nx}=-w_{Nx}^{(n+1)}\frac{\eta^{n+1}}{(n+1)!}
-\sum_{k=0}^{n-1}
\left (
w_{kx}^{(n-k)}\frac{\eta^{n-k}}{(n-k)!}
+u_{kz}^{(n-k-1)}\frac{\eta^{n-k-1}}{(n-k-1)!}
\right )
+\sum_{p=0}^{[(n-k-2)/3]} 2(-1)^{p}
\eta_z^{2p+1}
\]
\begin{equation}
\times \sum_{k=0}^{n-3p-2}
\left [
\left (
w_{kz}^{(n-k-3p-2)}
-u_{kx}^{(n-k-3p-2)}
\right )
\frac{\eta^{n-k-3p-2}}{(n-k-3p-2)!}
\right ]
~~~ (x=1),
\label{cccwba}
\end{equation}
\narrowtext
\noindent and
\begin{equation}  
w_n=0 ~~~ (x=0).
\label{cccw0a}
\end{equation}
Finally, after the solution $w_n$ is known, the system 
\begin{equation}  
\tilde{u}_{nx}=-\sum_{k=0}^{n-1} \left (u_{kx}
+w_{kz}\right )-w_{nz}-\tilde{w}_{(n+1)z}
\label{cccu}
\end{equation}
and 
\begin{equation}  
\tilde{u}_n=0 ~~~ (x=0)
\label{cccub}
\end{equation}
leads to the problem for $u_n$, 
\begin{equation}  
u_{nx}=-w_{nz}
\label{cccua}
\end{equation}
and 
\begin{equation}  
u_n=0 ~~~ (x=0),
\label{cccuba}
\end{equation}
solution of which completes the $n$th iteration step. In view of the exact
equations for $\widetilde{p_{n-1}}$, $\widetilde{w_n}$, and $\widetilde{u_n}%
, $ it is easy to justify by mathematical induction the equations for the
approximates $p_{n-1}$, $w_n$, and which are the principal objects of our
consideration here.

Since each of these problems is a linear (nonhomogeneous) ODE with linear
boundary condition, its solution is the sum of contributions from each term
of the RHS of the equation (which contribution is the solution of the ODE
with all other RHS terms put to zero and with homogeneous BC) and of each
term of the free-surface BC (the latter contribution is the solution of the
homogeneous ODE with the BC truncated to that single term). With this, it
can be seen, also inductively, that the order of magnitude of each term of $%
w_n$ is the product of certain powers of our four independent small
parameters---$A$, $1/L^2$, $R/L$, and $R/T$---with the sum of the exponents
being ($n+1$), the same for all terms of $u_n$ (we will say that the term
has the ``order of smallness'' $n+1$); and, moreover, all possible types of
terms of order $(n+1)$ are present in general, except for the terms of the
type (i.e. order of magnitude) $A^{n+1}$. Essentially the same orders of
magnitude are found in $u_n$, since, from (\ref{cccua}), it is the sum of
contributions of each term of $w_n$ (differentiated in $z$, which gives the
additional factor $1/L$ in the order of magnitude, and integrated in $x$,
which does not affect the order of magnitude at all). Similarly, $Rp_{n-1}$
is made of all possible terms of the order $n$ times the factor $1/L$ (in
this case, without any exceptions, since the terms of type $A^n/L$ \textit{%
are} present). One can note that the first three terms of the RHS (\ref
{cccwa}) give contributions which are smaller than $w_{n-1}$ by the factors $%
1/L^2$, $R/L$, and $R/T$, respectively, while the BC term $w_{(n-1)xx}\eta $
contributes terms which are smaller by the factor $A$ (including the terms
whose types contain $A^n$, such as $A^nR/L$). A similar role for $Rp_{n-1}$
is played by the BC term $Rp_{n-2}\eta $.

With the above order-of-magnitude structure of $u_n$ being known, it is easy
to see that each term of it is much smaller than some term of $u_{n-1}$, by
a small parameter $1/L^2$, $R/L$, $R/T$, or $A$. [It may seem at first sight
that terms of type, say, $A^n(R/L)$ present a difficulty: there is no $A^n$
in $u_{n-1}$ (we omit the common $1/L$). However, $%
A^n(R/L)=A[A^{n-1}(R/L)]$, and the type $A^{n-1}(R/L)$ \textit{is} present
in $u_{n-1}$.] It follows that 
\[
u_n\ll u_{n-1}
\]
Thus, the iterative corrections of $u$ are monotonously decreasing. (The
same can be seen for the sequences $\{w_n\}$ and $\{p_n\}$.)

All $u_n$ are substituted into the kinematic equation, and the
time-derivatives are eliminated by 
iterating the EE (similar to sec. \ref{s5}).
Then we have, in addition to the terms of the GEE (\ref{eee10})
(which are of
order two), also linear and nonlinear terms of order three and more (not
counting the factor $1/L$). Of these, the nonlinear dissipative (i.e. with
the total number of $z$-differentiations being even) terms are of a type
containing $A$ to the power at least two, and containing $R/L$ to the power
not less than one (since there are at least two differentiations). Thus, the 
$O_M$-type of each of these terms contains all the type-factors of the
(nondegenerate) nonlinear dissipative term $\propto R(\eta ^2)_{zz}$, $\sim
(1/L)A^2(R/L)$, and, except for this principal nonlinear dissipative term,
also some additional small factor, the product of powers of the small
parameters $A$, $1/L^2$, and $R/L$, which makes that term negligible.
Therefore, we do not have to retain any nonlinear dissipative term other
than the principal one (which is of order three). Similarly, of all the
dispersive terms (both linear and nonlinear), only the (nondegenerate,
second-order) term $\propto \eta _{zzz}$ is retained.

As we noted in the main text, all the linear dissipative terms could be
similarly neglected as compared with the linear second-derivative one, but
the latter is degenerate if $\delta \ll R$. Therefore, we should consider
these linear terms. If the type of a linear term does not contain a factor
which is a power of $R/L$, that term is just a $(2n+1)$-derivative one,
clearly a dispersive term. Thus, the dissipative linear terms we are
interested in must have the ($l$th-derivative) form $R^k\eta ^{(l)}$, which
leads immediately to Eq. (\ref{eee25}). 
\bibliographystyle{prsty}

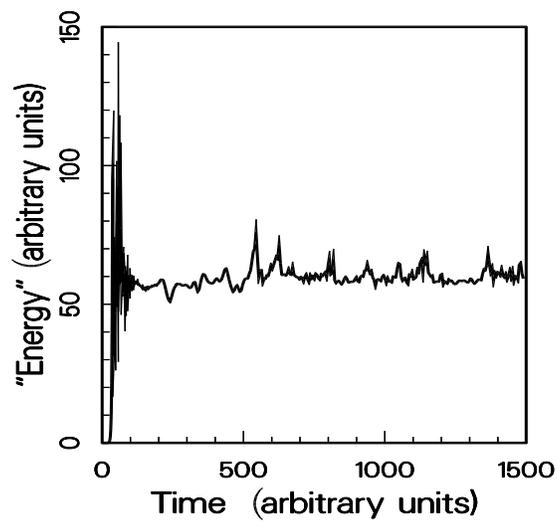
\begin{figure}
\caption{
Evolution of energy illustrating that the
solutions of Eq. (\protect\ref{eee12}) remain bounded.
}
\label{f1}
\end{figure}
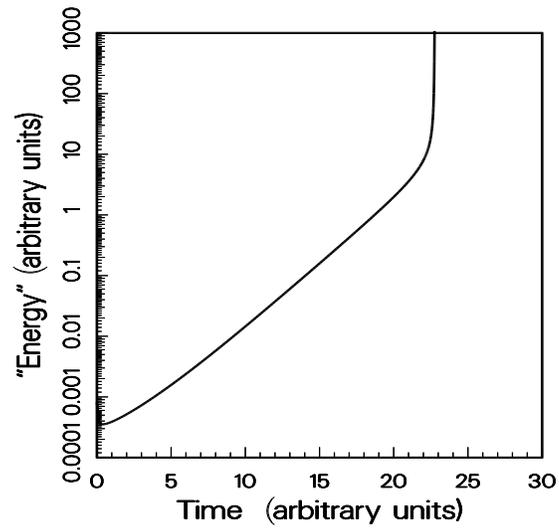
\begin{figure}
\caption{
Evolution of energy indicates that the
solutions of Eq. (\protect\ref{eee12n}) blow up.
}
\label{f2}
\end{figure}
\end{document}